%Paper: cond-mat/9503172
%From: fendley@voltaire.usc.edu (Paul Fendley)
%Date: Fri, 31 Mar 95 17:19:50 -0800
%Date (revised): Wed, 13 Sep 95 14:13:30 -0700
%Date (revised): Wed, 13 Sep 95 15:00:31 -0700

% This paper uses harvmac macros, which are available from the
% cond-mat bulletin board via `get harvmac.tex' in the subject line.
% If using harvmac for the first time, be sure to set the
% site-specific options for printers at the beginning
%
% This paper also has two uuencoded figures, which should arrive
% in a separate file. Instructions for unpacking them are at
% the beginning of that file.

\input harvmac
\lref\GZ{S. Ghoshal and A.B. Zamolodchikov, Int. J. Mod.
Phys. A9 (1994) 3841, hep-th/9306002.}
\lref\Zamo{Al.B. Zamolodchikov, Phys. Lett. B253 (1991) 391.}
\lref\YY{C.N. Yang and C.P. Yang, J. Math.  Phys. 10 (1969)
1115}
\lref\Zamtba{Al.B. Zamolodchikov, Nucl. Phys. B342 (1991) 695.}
\lref\KM{T.R. Klassen and E. Melzer, Nucl. Phys. B338 (1990) 485.}
\lref\FSW{P. Fendley, H. Saleur and N.P. Warner,
Nucl. Phys. B430 (1994) 577, hep-th/9406125.}
\lref\Stone{M. Stone and M.P.A. Fisher, Int. J. Mod. Phys.
B8 (1994) 2539, cond-mat/9402040.}
\lref\ZZ{A.B. Zamolodchikov and Al.B. Zamolodchikov, Ann. Phys. 120
(1979) 253.}
\lref\Wen{X.G. Wen, Phys. Rev. B41 (1990) 12838;
Phys. Rev. B43 (1991) 11025. }
\lref\KF{C.L. Kane and M.P.A. Fisher, Phys. Rev. B46 (1992) 15233.}
\lref\Moon{K. Moon, H. Yi,  C.L. Kane,
S.M. Girvin and M.P.A. Fisher, Phys. Rev. Lett. 71 (1993) 4381.}
\lref\CW{C. De C.Chamon and X.G. Wen, Phys. Rev. Lett.  70 (1993)
2605.}
\lref\FN{A. Furusaki and N. Nagaosa, Phys. Rev. B47 (1993) 3827.}
\lref\ALxray {I. Affleck and A.W.W. Ludwig, J. Phys. A27 (1994) 5375,
cond-mat/9405057.}
\lref\boson{A. Luther and I. Peschel, Phys. Rev. B12 (1975) 3908;
\hfill\break S. Coleman, Phys. Rev. D11 (1975) 2088.}
\lref\KB{L.P. Kadanoff and A.C. Brown, Ann. Phys. 121 (1979) 318.}
\lref\Hald{F.D.M. Haldane, J. Phys. C14 (1981) 2585.}
\lref\exper{F.P. Milliken, C.P. Umbach and R.A. Webb,
``Indications of a Luttinger Liquid in the Fractional Quantum Hall
Regime'', to appear in Solid State Communications.}
\lref\WA{E. Wong and I. Affleck, Nucl. Phys. B417 (1994) 403,
cond-mat/9311040.}
\lref\Guin{F. Guinea, Phys. Rev. B32 (1985) 7518.}
\lref\FLSi{P. Fendley, A. Ludwig and H. Saleur,
Phys. Rev. Lett. 74 (1995) 3005, cond-mat/9408068.}
\lref\FSZII{P. Fendley, H. Saleur and Al.B. Zamolodchikov,
Int. J. Mod. Phys. A8 (1993) 5751.}
\lref\FLSjack{P. Fendley, F. Lesage and H. Saleur,
J. Stat. Phys. 79 (1995) 799, hep-th/9409176.}
\lref\callan{C.G. Callan and I.R. Klebanov, Phys. Rev. Lett. 72
(1994)
1968,  hep-th/9311092; C.G. Callan, I. Klebanov, A.W.W. Ludwig and
 J. Maldacena, Nucl. Phys. B422 (1994) 417,
 hep-th/9402113; J. Polchinski and L. Thorlacius,
Phys. Rev. D50 (1994) 622, hep-th/9404008.}
\lref\TS{M. Takahashi and M. Suzuki, Prog. Th. Phys. 48 (1972) 2187.}
\lref\AFL{N. Andrei, K. Furuya, and J. Lowenstein, Rev. Mod. Phys. 55
(1983) 331; \hfill\break
A.M. Tsvelick and P.B. Wiegmann, Adv. Phys. 32 (1983) 453.}
\lref\FI{P. Fendley and K. Intriligator, Nucl. Phys. 372 (1992)
533.}
\lref\JNW{G. Japaridze, A. Nersesyan and P. Wiegmann, Nucl. Phys.
B230 (1984) 511.}
\lref\RS{N.Yu. Reshetikhin and H. Saleur, Nucl. Phys. B419 (1994)
507.}
\lref\FT{L.D. Faddeev and L.A. Takhtajan, Phys. Lett. A85 (1981)
375.}
\lref\KFii{C. Kane and M.P.A. Fisher,  Phys. Rev. B 46 (1992) 7268.}
\lref\AR{W. Apel and T.M. Rice, Phys. Rev. B26 (1982) 7063.}
\lref\ALkond{I. Affleck and A.W.W. Ludwig, Nucl. Phys. B360 (1991)
641.}
\lref\AffLH{for a review of this method, see I. Affleck in Les
Houches 1988 {\it Fields, strings, and critical phenomena}, ed. by E.
Br\'ezin and J. Zinn-Justin (North-Holland, 1990).}
\lref\Poly{A.M. Polyakov, Phys. Lett. 72B (1977) 224.}
\lref\Haldexc{F.D.M. Haldane, Phys. Rev. Lett. 67 (1991) 937.}
\lref\Guineanew{F. Guinea, G. Gomez Santos, M. Sassetti and
M. Ueda,  ``Asymptotic tunneling conductance in Luttinger
liquids'', cond-mat/9411130.}
\lref\Weissi{U. Weiss, M. Sassetti, T. Negele
and M. Wollensak, Z. Phys. B 84 (1991) 471.}
\lref\Weiss{U. Weiss, R. Egger and M. Sassetti,
``Low-temperature nonequilibrium
transport in a Luttinger liquid'', to appear in Phys. Rev. B.}
\lref\Tsv{A.M. Tsvelick, ``Capacitance and transport through an
isolated impurity in one-dimensional Luttinger liquid'',
cond-mat/9409027.}
\lref\Sch{A. Schmid, Phys. Rev. Lett. 51 (1983) 1506.}
\lref\FZ{M.P.A. Fisher and W. Zwerger, Phys. Rev. B32 (1985) 6190.}
\lref\GHM{F. Guinea, V. Hakim and A. Muramatsu, Phys. Rev. Lett. 54
(1985) 263.}
\lref\Zisn{Al. B. Zamolodchikov, Nucl. Phys. B358 (1991) 497.}
\lref\MW{B.M. McCoy and T.T. Wu, Nuovo Cimento 56B (1968) 311.}

\def\t{\theta}
\def\ep{\epsilon}
\def\l{\lambda}
\def\<{\langle}
\def\>{\rangle}
\noblackbox

\Title{\vbox{\baselineskip12pt
\hbox{cond-mat/9503172, USC-95/007}}}
{\vbox{
\centerline{ Exact non-equilibrium transport through }
 \vskip 4pt
\centerline{ point contacts in quantum wires and}
 \vskip 4pt
\centerline{ fractional quantum Hall devices}
}}
\centerline{P. Fendley$^1$, A.W.W. Ludwig$^{2 \dagger} $ and H.
Saleur$^{1*}$ }
\bigskip
\centerline{$^1$ Department of Physics,
University of Southern California}
\centerline{Los Angeles CA 90089-0484}
\medskip
\centerline{$^2$ Department of Physics,
University of California}
\centerline{Santa Barbara CA 93106}
\vskip.3in

We have recently
calculated exact non-equilibrium quantum
transport properties through a point contact in
a Luttinger liquid.
Using a particular quasiparticle basis
of the Hilbert space dictated by integrability,
we here compute explicitly the exact $I(V)$
characteristic and conductance
{\it out of equilibrium} as a function of driving
voltage $V$ and
temperature $T$. These are described by
 universal scaling functions of two variables,  the scaled
point-contact
interaction strength, and $V/T$.
The differential-conductance curve as a function of
the interaction strength
broadens significantly  as $V/T$ is increased,
and develops a pronounced  maximum at a (universal) critical
value $(e V/k_B T)=7.18868\dots$. In addition, we derive
an exact duality between strong and weak backscattering.
 The theory presented here has recently been
realized experimentally in  resonant
 tunneling-transport experiments between
edge states in fractional quantum Hall effect devices.
In this context the exact duality is
between electron tunneling and Laughlin-quasiparticle tunneling.

\bigskip
\bigskip
\noindent $^{\dagger}$  A.P. Sloan Fellow

\noindent $^*$ Packard Fellow

\Date{3/95}

\newsec{Introduction}

Non-equilibrium quantum transport in fully-interacting systems is
a barely-explored territory of theoretical physics.
Equilibrium statistical mechanics of interacting
systems, on the other hand, can sometimes be studied reliably by
using
powerful  field-theoretical techniques
(including conformal  field theory and the Bethe ansatz).
Bethe-ansatz integrability  is useful in equilibrium  even
in the absence of scale invariance,
permitting one  to study the
exact crossover  behavior between critical points.
 However, integrability has for the most
part been useful for calculating only
thermodynamic quantities, excluding correlation
functions  and transport at non-zero temperature.
However, by using the ``quasiparticle approach''
to integrability, we have recently shown that,
quite unexpectedly, one {\it  can}
compute exact transport properties through a point contact,
even out of equilibrium \FLSi.
Here we discuss non-equilibrium
transport through point contacts  in a Luttinger
liquid. This model is realized in resonant tunneling experiments
through point contacts in quantum Hall effect devices \exper.

 The key observation in  \FLSi\ was that (i) tunneling
is integrable and  that (ii)  integrability
defines a quasiparticle
basis of the Hilbert space of the leads which is
particularly suited to computing transport. In this basis,
scattering
processes at the point contact proceed without
particle production. These quasiparticles are not
free.  However, their interactions can be incorporated
exactly
into {\it non-fermi distribution functions}, which
govern the  filling of  single-particle levels (orbitals)
with quasiparticles in the thermodynamic limit.
Even though these  distribution functions are not that of
Fermi-Dirac, in an  integrable model they
can be computed exactly using thermodynamic
Bethe ansatz (TBA) technology.
Once this quasiparticle basis arising
from integrability has been identified, non-equilibrium
transport properties such as the exact $I(V)$ curve and
the conductance through the point contact can be
computed exactly using a kinetic (Boltzmann) equation
for these quasiparticles.
This is possible even though the point-contact
interaction is non-Gaussian, because the constraints of
integrability give the exact $S$ matrix of transmission and
reflection
amplitudes for these quasiparticles
scattering off the point contact. This
elastic single-quasiparticle $S$ matrix
is momentum-dependent, and,  since there is no
quasiparticle production in this basis, it
is unitary. The S-matrix for scattering of
a multi-particle state off the contact
factorizes into a product of
single particle S-matrices.
  This means that
 reflection and transmission processes
of  single  quasiparticles by the point contact
occur successively (``one-by-one''),  and
are the only scattering events.  Therefore,  these single
quasiparticle processes describe the effect of the
point-contact interaction fully and exactly.

The purpose of this paper is threefold:
 (i) We give the details of our
exact non-equilibrium transport calculations through
point contacts, using Bethe ansatz integrability.
 We have recently reported
briefly on some  of our results in \FLSi. (ii): We give
universal explicit curves for the $I(V)$ characteristic
and the conductance of those point contact devices,
for all temperatures,
for comparison with future experiments.
(iii): We prove the existence of  an exact duality symmetry in this
interacting system between weak and strong backscattering.
In the context of the quantum Hall effect
this corresponds to a duality between electron tunneling
and Laughlin-quasiparticle tunneling.

The Luttinger model is one of the
simplest non-fermi-liquid metals \Hald. It consists
of left- and right-moving gapless excitations
at the two fermi points in an interacting 1-dimensional
electron gas.  In the past, this model had been difficult to
realize experimentally. This is simply because
in a one-dimensional conductor (such
as  a quasi-one-dimensional quantum
wire so thin that the transverse modes
are frozen out at low temperature), random impurities
occur in the fabrication process. These impurities lead to
localization due to backscattering processes
between the excitations at the two fermi points.
In other words, the random impurities generate a mass
gap for the fermions.
However, the edge excitations
at the boundary of samples prepared in  a fractional
quantum Hall state should be  extremely  clean realizations of
the Luttinger non-fermi liquids \Wen. These are stable because
 for $1/\nu$ an odd integer, the excitations only move in one
direction on a given edge.  Since the right and left edges are
far apart from each
other, backscattering processes
due to random impurities in the bulk cannot localize  those
extended edge states. Moreover, the Luttinger interaction
parameter is universally related  to the filling fraction
$\nu $ of the quantum Hall state in the bulk sample
by a topological argument based on the underlying Chern-Simons
theory, and does therefore not renormalize.
The edge states should therefore provide an extremely
clean  experimental realization of the Luttinger model.

We study the tunneling conductance through
a local backscattering potential in the
 Luttinger model, which gives a fingerprint
of the non-fermi liquid state in the  Luttinger metal.
 This  situation is realized in
resonant tunneling experiments
through a point contact in $\nu =1/3$
quantum Hall devices  \refs{\Wen,\KF,\Moon}.
The point contact causes backscattering  between right-
and left-moving edge excitations,
 but since the coupling is
only at a point, it can be controlled experimentally
(and theoretically). The tunneling conductance
can thus be viewed  as  a spectroscopy of the Luttinger
non-fermi-liquid
state in the quantum Hall edges.
Indeed,
the experimentally-measured linear-response conductance
agrees remarkably well
with our exact predictions based on the Luttinger model \FLSi.
This appears to provide  very convincing evidence
that the Luttinger model
describes the  edge state in fractional quantum Hall
devices.

In this paper we study non-equilibrium transport
through such an impurity in detail.
Studying transport properties is crucial for making contact
with experiment;  thermodynamic properties such as
the  specific heat arising from a point contact
are clearly not accessible experimentally.
 In the quantum  Hall experiments one uses
a 4-terminal geometry of the quantum Hall bar,
which is long in the $x$-direction and short in the $y$-direction.
The left-moving (upper)  edge of the Hall bar
 is connected to battery on the right
such that the charge carriers are injected into
the left-moving  lead of the
Hall bar with an equilibrium thermal
distribution at chemical potential $\mu_L$. Similarly,
the right-moving carriers  (propagating in the lower edge)
are injected from the left, with a thermal distribution
at chemical potential $\mu_R$. The difference
of chemical potentials of the injected
charge carriers is the driving voltage $V= \mu_L-\mu_R$.
If $V>0$, there are more carriers injected from
the right than from the left, and a ``source-drain''
current flows from the right to the left,
 along the $x$-direction of the Hall bar.
In the absence of the point contact,
the driving voltage places the right and left
edges at different potentials (in the $y$-direction,
perpendicular to the current flow),  implying that
the ratio of source-drain
current to the driving voltage $V$ is the Hall conductance.

In the absence of any point-contact interaction, the
source-drain current $I_0(V)$ may be computed
in a variety of ways
(see e.g.\ \refs{\AR,\KF}). The resulting conductance
is $\nu e^2/h$ (in linear response and at finite driving voltage
$V$).
When we include a point-contact interaction
at finite driving voltage,  more
of the left-moving carriers injected from the
right are backscattered than those injected
from the left, resulting in a loss of
charge carriers from the source-drain current.
In this case we write the total source-drain current
as $I(V) = I_0(V) + I_B(V)$, where $I_B(V)$ is
the (negative) backscattering current, quantifying
the loss of current due to backscattering at the point
contact.
It is this backscattering current which we compute
exactly here, for finite driving voltage $V$ and
all temperatures $T$. This computation is possible  since
backscattering does not
deplete the (infinite) right and left reservoirs
(the battery), so that
we can use the individual thermal distribution functions
for the left and right reservoirs.

The $I(V)$ characteristic as well as the conductance
are described by three parameters, $V$, $T$ and
$T_B$, where $T_B$ (analogous to the Kondo temperature in the Kondo
effect)
is  a measure of the point contact interaction strength. However,
due to the underlying
quantum critical point, these  observables
are described by universal scaling functions
of two ratios, the scaled interaction strength
$T_B/T$,  and $V/T$. We display
the exact differential conductance
$G(T_B/V,V/T) \equiv {\partial I(V)\over \partial V}$
in fig.\ 1.
As apparent from fig.\ 1,
an interesting  prediction of our exact solution
is that the differential
conductance
exhibits a pronounced maximum as a function of the
point contact interaction strength,
  whenever the ratio
$V/T$ exceeds a critical value (which depends on the filling $\nu$,
i.e.\ on the Luttinger interaction constant). This
is a pure non-equilibrium effect, since the maximum
occurs only at finite driving $V\not = 0$.
Of course, the total current still {\it de}creases
as the point-contact  interaction strength is increased.
This feature is consistent with the first-order perturbative results
\KF\ for the $I(V)$ curves
at zero temperature, in the strong and weak backscattering
limits.

When
there is a single relevant operator corresponding to the
impurity, the model is integrable, so
we compute the current and conductance exactly.
There is only one relevant operator when $\nu=1/3$, so
integrability is generically observed without any fine tuning. Any
other sample-specific details appear only in the irrelevant
operators.
Integrability
allows the definition of a basis of massless
charge-carrying `quasiparticle' excitations \refs{\GZ,\FSW}.
These quasiparticles are scattered one-by-one
off the impurity with a momentum-dependent
one-particle scattering matrix $S$
of transmission and reflection amplitudes. These amplitudes
can be determined exactly by imposing the constraints of
integrability, including the boundary Yang-Baxter equation
and the boundary-crossing relation \GZ.
Furthermore, the quasiparticles are characterized by an
thermal distribution
function which can be calculated exactly using the
thermodynamic Bethe ansatz \refs{\YY,\Zamtba}.
This special behavior is
a consequence of integrability, and it
allows us to derive an exact rate (Boltzmann)
equation for
the conductance in this interacting theory.

We should note that many of these results have been checked by
using a completely different method of computation. Instead of the
non-perturbative methods to be described below, one can study the
model
perturbatively in the interaction strength $T_B$. Closed-form
results for all the perturbative coefficients of the free energy
have been found; a simple technical assumption then also leads to
closed-form expressions for all the perturbative
coefficients of the
conductance, even at finite $V$ \FLSjack.
There is complete qualitative and quantitative
agreement between the two approaches.

The outline of this paper is as follows:
In sect.\ 2 we discuss how this system is
realized in experiments on resonant tunneling
through point contacts in fractional quantum Hall devices.
In sect.\ 3, we map the problem into two decoupled theories,
one of which is affected by the backscattering interaction,
and another one which is not. In sect.\ 4 we find the exact
quasiparticle spectrum, exact $S$ matrices for quasiparticles
scattering among themselves and off of the boundary, and
the exact thermal distribution functions, using the fact that
the interacting theory is integrable.
In sect.\ 5, we derive an exact (Boltzmann) equation for the
current in terms of the $S$ matrix and the distribution functions.
This gives us equations for the exact conductance. These
can easily be solved numerically, and we present curves for
a variety of values. At zero temperature, the equations simplify
and we present more explicit analytic results in
sect.\ 6. This enables us to derive an exact duality between
strong and weak backscattering limits, out of equilibrium.

\newsec{The fractional quantum Hall effect}

\subsec{Experimental setting for resonant tunneling}

Experiments on resonant tunneling between two
$\nu=1/3$ edges have recently been performed by Milliken,
Umbach and Webb \exper.
We briefly review the experimental setup schematically
(for details see \refs{\exper,\Moon}).
A fractional quantum Hall state  with filling fraction $\nu=1/3$
is prepared in the
bulk of a quantum Hall bar (discussed in Section 1).
This means that the bulk quantum Hall state
is prepared  in a Hall insulator state
(longitudinal conductivity $\sigma_{xx}=0$),
and that the (bulk) Hall
resistivity  is on the $\nu=1/3$ plateau
where $\sigma_{xy}=(1/3)e^2/h $.  This is achieved
by adjusting the applied magnetic field, perpendicular to
the plane of the bar. Since the plateau is broad, the applied
magnetic field can be varied over a significant range without
affecting the filling of $\nu=1/3$.
Next, a gate voltage $V_g$ is applied perpendicular to
the long side of the bar, i.e. in the $y$ direction (see Section 1)
at $x=0$. This has the effect of bringing the right and left
moving edges close to each other near $x=0$, forming
a point contact. Away from the contact there is no backscattering
(i.e.\ no tunneling of charge carriers)
because the edges are widely separated, but now
charge carriers can hop from one edge to
the other at the point contact.

The linear-response source-drain conductance as
a function of temperature and gate voltage $V_g$
has been measured experimentally \exper.
As the gate voltage is swept through, the conductance
signal shows a number of resonance peaks, which
sharpen as the temperature is lowered.
These resonance peaks occur for particular values
$V_g = V_g^*$ of the gate voltage, due to tunneling through
localized states in the vicinity of the point contact.
Ideally, on resonance, the source-drain conductance
is equal to the Hall conductance without point contact,
i.e. $G_{resonance} = (1/3) e^2/h$. This
value is independent of temperature, on resonance.
Now, measuring the linear response conductance
as a function of the gate voltage near the
resonance, i.e. as a function of
$ \delta V_g \equiv V_g-V_g^* $, at a number of different
temperatures $T$, one gets resonance curves, one for each
temperature.
These peak at $\delta V_g =0$. Scaling arguments imply
\KF\ that those experimental
conductance  curves should collapse
onto a single universal curve when plotted
as a function of $\delta V_g/T^{2/3}$. Indeed they
do collapse quite well \exper. Moreover, the resulting
universal curve should be a unique signature and fingerprint
of the $\nu =1/3$ edge state in the leads connected to the
point contact.

\subsec{ Comparison of  the linear response resonance
lineshape with  our exact Luttinger model predictions}

To make contact with the Luttinger model, we  state (as will be
explained below in more detail)
that the experimental parameter $\delta V_g$ should
be related to the Luttinger backscattering interaction
by $\delta V_g \propto T_B^{2/3} $. In particular, at the resonance
value $T_B=0$, there is no backscattering at all.

We have compared \FLSi\ our exact
predictions for the linear response conductance
scaling curve with the experimental data \exper\ as well as
Monte Carlo calculations in \Moon.
The agreement
between the Monte Carlo simulation and our exact scaling curve is
excellent.  The exact value of the universal parameter $K$ (defined
so that $G(X)=KX^{-6}$ for $X$ large and $G(X)=(1-X^2)/3$ for $X$
small)
is $K=3.3546...$ (where $X\approx .74313 (T_B/T)^{2/3}$).  (The value
$K\approx 2.6$ quoted in
\Moon\ seems to have been slightly underestimated there.)

 The comparison with the existing  experiments
by Milliken et al is not completely straightforward, since
the conductance at the resonance peak in the
experimental data decreases with
temperature and is well below its resonance value $ e^2/3h$.  This
difficulty arises since in order to achieve the resonance condition
in
the Luttinger model, two parameters need to be tuned,
since the point contact  will in general not possess
reflection-parity symmetry  about the point $x=0$ \KFii.
 In the experiments performed
so far only one parameter, namely $V_g$,  has been tuned.
For that reason, the conductance peaks do
not have their maximum height
 $G_{resonance} = (1/3) e^2/h$, but are smaller, and, furthermore,
the peak height does decrease with temperature,
reflecting the fact that the experimental peak
is not a perfect resonance.
This problem can be remedied in a future experiment, by
varying {\it two} parameters, namely $V_g$ {\it and} the magnetic
field on the plateau, instead of only one parameter,
to achieve resonance.
Nevertheless, even when only the
gate voltage is tuned to resonance, the experimental data
for the conductance signal as a function of the gate
voltage and temperature collapse well onto single
scaling curve. Thus it makes sense to compare the
this experimental curve with  our exact
conductance curve, computed from the Luttinger model.
The agreement is quite good,
given the large scatter of the data in the tail of the
resonance curve.  In particular, the data clearly show the predicted
$G \propto T^4/ (V_g-V^*_g)^6$ behavior in the tail.

\subsec{ Predictions for future non-equilibrium transport
measurements}

In principle, there is no reason why the above mentioned
measurements could not be extended to finite driving
voltage $V$ [one should not confuse the driving voltage $V$,
which is the difference between the chemical potentials
between the injected left- and right-moving charge carriers,
with the gate voltage $V_g$, which gives rise to the
coupling constant $T_B$ in the Luttinger liquid theory].
So far, only the linear-response conductance,
$G = \lim_{V \to 0} \  I(V)/V$ has been measured. More generally,
one could attempt to  measure the conductance at {\it finite} driving
voltage $V$.
Note that our exact results predict the shape of the universal
scaling function
$$ G(T_B/T,V/T), \qquad T_B = C (\delta V_g)^{3/2} $$
as  a function of {\it two } ratios. The non-universal parameter
$C$ is determined by fitting the experimental data to the universal
curve; this is the only unknown quantity. The linear-response
conductance is the limit of this function as $V\to 0$,
where it becomes a function only of one ratio $T_B/T$.
Note that the conductance at finite driving voltage $V$
also describes a resonance lineshape.
A particularly interesting feature  of the
conductance at finite driving voltage $V$ is seen in fig.\ 1.
The conductance as a function
of $T_B \propto (\delta V_g)^{3/2} $
develops a pronounced peak when the
ratio of driving voltage to temperature $e V/k_B T$ exceeds
a critical value  $7.188...$.  For a typical low temperature
$T=50mK $  used in the data of \exper,
 this would correspond ($10 k_B K = 1meV$,  or, ${5 \mu V \over
50mK} = 1 $)
to a driving voltage of $V^* \approx  35 \mu V$.
The current data were taken at an `excitation voltage' of $1 \mu V$,
which corresponds to a ratio $e V/k_B T \approx 0.2$.  We have
plotted the exact results for the aforementioned scaling functions
in Figs.\ 1 and 2.
One sees clearly from these plots that  $eV/k_B T\approx 0.2$
corresponds to the linear response regime.
Perhaps the most significant feature displayed in
Fig.\ 2 is a very dramatic {\it non-equilibrium broadening}
of the resonance curve
for values of the ratio $V/T$ even well  below the occurance
of the maximum. In terms of numbers, the curve broadens
by a dramatic amount already at $eV/k_B T \approx 2$ or $ 3$,
well before the onset of the maximum (at $eV/k_B T \approx  7.188$).
This broadening should be easily visible experimentally,
since it would correspond to an excitation voltage
$V \approx 15 \mu V$, just a factor of 15 higher than
the ones used by Milliken et al., at $T= 50mK $.

Clearly, at larger voltages one will have to worry about
larger currents flowing in the sample, which might lead to
possible complications.
These complications would for example
include non-universal effects arising
from sample heating which are clearly
not included in the Luttinger theory
and which could mask the underlying universal
Luttinger non-fermi liquid physics. However
one should notice that for most part of the conductance
curve the conductance is very small, implying
flow of very small currents, so that sample
heating should not appear to be a problem.

Futhermore,
it is important to notice
that the scaling function depends only on the
ratio $V/T$. This means the same critical ratio $(V/T)^*= 7.188$
could also be achieved by working at lower temperatures.
It might be reasonable experimentally to perform
measurements at lower temperatures. This would reduce
even more the values of  the driving voltage where
the onset of the universal non-equilibrium features
(the most dramatic one being the broadening)
become clearly visible.

\newsec{Bosonizing and mapping to an integrable model}

The Luttinger model is the most general model of a single massless
fermion in one dimension. The fermion interaction is governed
by a single parameter $g_{Lutt}$, so that when $g_{Lutt}>0$
the interaction is repulsive and $g_{Lutt}<0$ it is attractive.
In the absence of the
impurity, the Hamiltonian is
\eqn\LuttHam{H=\pi\int_{-l}^l dx
\left(J_L^2+J_R^2+ g_{Lutt} J_LJ_R\right)}
where the left movers are coupled to the right.
We use the well-known map of this model to
free massless left and right moving bosons $\Phi_L$ and
$\Phi_R$ \boson. To do so one bosonizes the currents
 $J_L=-{1\over 2\pi}\partial_x\Phi_L$ and $J_R={1\over
2\pi} \partial_x\Phi_R$. The fermion operators take also the form
$\Psi_R=:e^{i\Phi_R}:$ and $\Psi_L=:e^{i\Phi_L}$:.
By introducing new fields \AffLH\
 using the transformation (canonical up to a global factor
$\sqrt{\alpha}$)
\eqn\newfields{\eqalign{\Phi_R=&{\alpha+1\over 2}
\varphi_R+{\alpha-1\over
2}\varphi_L\cr
\Phi_L=&{\alpha-1\over 2}\varphi_R
+{\alpha+1\over 2}\varphi_L,\cr}}
one can decouple the interaction in the bosonized hamiltonian that
reads then
\eqn\newhamil{H_0={\pi\over\nu} \int_{-l}^l dx
\left[j_L^2+j_R^2\right]=
{1\over 8\pi\nu}\int_{-l}^l \Pi^2+
\left(\partial_x\varphi\right)^2,}
where $\Pi $ is canonical momentum conjugate to $\phi$.
The coefficient $\nu $ is
related to the old Luttinger coupling via
$g_{Lutt}=-2{\alpha^2-1\over \alpha^2+1}$,
${1\over\nu}={2\alpha^2\over 1+\alpha^2}$.
The currents
 $j_L=-{1\over 2\pi}\partial_x\varphi_L$ and $j_R={1\over 2\pi}
\partial_x\varphi_R$  are then normalized so that
$ \< j_L(x_1)j_L(x_2)\> = {\nu \over (2\pi)^2 (x_1-x_2)^2}$
which makes  the meaning of $\nu$ as a chiral $U(1)$ anomaly
explicit.
  The operators
$\psi_R=:e^{ (i/\nu)\varphi_R}:$ and
$\psi_L=:e^{ (i/\nu)\varphi_L}:$ are sometimes
referred to as  ``Luttinger hyperfermion''
 operators when $1/\nu$ is an odd integer \Stone.
There are two $U(1)$ symmetries of this model
corresponding to left and right (hyper) fermion number: we introduce
the corresponding conserved charges
$ Q_L= \int_{-l}^ldx j_L$, and
$ Q_R= \int_{-l}^ldx j_R$.
 The corresponding ``hyperfermions'' have charge $e$ as do the
original
electrons, as can be seen by computing their commutators
with the charge operators.

We now include an impurity, which we assume is localized at the
origin
$x=0$. It couples left- and right-moving fermions with the
interaction
$\Psi^\dagger_L(0) \Psi_R(0) + \Psi^\dagger_R(0) \Psi_L(0)$.
Such a coupling destroys the separate conservation
of the two charges $Q_L$ and $Q_R$: only the  {\it total} charge
 $ Q_L+Q_R$ remains conserved, corresponding to
simultaneous phase translations of $\varphi_L$ and $\varphi_R$.
In a renormalized effective
theory, all terms that are not forbidden by total
charge conservation symmetry can appear in the hamiltonian.
These may be
represented in terms of the bosons $\varphi_L,\varphi_R$ by a
backscattering contribution to the Hamiltonian \refs{\Wen,\KF} :
\eqn \delH {
H_B ={1\over 2}\sum_n \lambda_n
\{ e^{i n\varphi_L(x=0)} e^{-i n\varphi_R(x=0)}
+e^{-i n\varphi_L(x=0)} e^{i n\varphi_R(x=0)} \}.  }
When we describe the Luttinger theory in terms
of the bosons $\varphi_L$ and $\varphi_R$, i.e.\
after the Luttinger bulk interaction $g_{Lutt}$
has been disentangled, this term
describes hopping of quasiparticles
of charge $\nu e$
(the Laughlin quasiparticles in the case of FQHE)
between the left- and right-moving edges.
 Only the terms with $n^2< 1/\nu$ are relevant.
For $\nu >1$ all terms are irrelevant; this case is
more appropriately described by a ``dual'' picture \KF. At
$\nu=1$ there is a single exactly marginal operator. This preserves
the conformal invariance of the fixed point, so conformal
techniques are applicable, making possible the calculation of
the complete
partition function \callan. We now focus on
the case $\nu <1$. For the model to be integrable,
it seems that only one relevant perturbation is allowed \GZ.
Thus when
$1>\nu>1/4$, the backscattering interaction
is {\it automatically} integrable
without {\it any} fine-tuning. The
experimentally-measured value of $\nu =1/3$
falls into this range of
parameters. For $\nu<1/4$, the model has to be fine-tuned
in order to be integrable. In particular, for
$  1/9 < \nu<1/4 $ a single parameter ($\lambda_2$)
 needs to be tuned  to zero in order to achieve integrability.
 In general, we will be treating the model with
$\lambda_n=0$ for $n\geq 2$.

In order to transport net charge through the impurity,
we place the injected  left movers and the right movers at different
chemical
potentials.
This amounts to adding a term $e(Q_L-Q_R)V/2$ to the Hamiltonian
describing the thermal weighting  of the injected charge carriers.
Notice that
even though this problem is out of equilibrium
for non-zero $V$, the charge carriers injected into
the leads are thermally weighted with an equilibrium
distribution function corresponding to $\mu_L$ or $\mu_R$. Thus we
are studying the coupling between two equilibrium distributions.

It is convenient to introduce the ``backscattering temperature''
$T_B$, which  is the scale generated by
the relevant  point contact
coupling constant at one point:  $$T_B=C_1
\lambda_1^{1/(1-\nu)}$$ where $C_1$ is some non-universal
constant.  When there is no backscattering, $T_B=0$.
The exponent follows from a simple perturbative
analysis \KF.
For $\nu< 1/2$ and $T>0$ there are actually
no divergences in the coefficients of the perturbative expansion;
we exploit this fact in Sect.\ 6 to
derive $C_1$ exactly.
The conductance $G$ and current $I$ are universal scaling
functions of two  dimensionless
ratios, for example  $T_B/T$ and $V/T$.
 Since $T_B$ is a scale introduced by
the (local)  impurity, a bulk quantity
like the leading part of the
thermal distribution functions is
independent of the point contact interaction and
can be computed without the point contact. In other
words, the corrections
vanish with the size of the system.

We now map the model including the impurity to an
integrable model. By taking (non-local) linear combinations
of the bosons, we can first map the left- and right-moving
bosons to a purely left-moving system. Calling these left movers
{\it even}
and {\it odd}, we have
\eqn\evenodd{\eqalign{
\phi^e(x+t) \equiv &
{1 \over \sqrt{2}} [ \varphi_L(x,t)+\varphi_R(-x,t)],
\cr
\phi^o(x+t)
\equiv &{1 \over \sqrt{2}} [ \varphi_L(x,t)-\varphi_R(-x,t)]\cr}}
where these particles are defined on $-l < x < l$. One replaces
$j_L,j_R$ with $j^e, j^o$ in the Hamiltonian \newhamil,
where $j^{e/o}(x+t)=
(1/\sqrt{2}) [ j_L(x,t) \pm j_R(-x,t)]$ are the charge densities of
even and odd bosons.  For purposes
of computing the backscattering current,
the non-locality of this transformation does not
matter (it clearly would if we were to compute
spatially-dependent Greens functions).
Two major simplifications arise from this
change of basis. First, the interaction involves only the combination
$
\varphi_L(x=0) - \varphi_R(x=0)$ so only the odd boson $\phi^o$
interacts,
while the even boson $\phi^e$ remains free \WA.  (This was the
crucial
step in solving the X-ray edge problem in a Luttinger liquid
\ALxray.)
Second, the backscattering current $I_B$ can be entirely expressed
in terms of the odd boson theory. This is easy to see: The even and
odd charges are related to the charges of the original left- and
right-moving edges by $\Delta Q = Q_L-Q_R =\sqrt{2} Q^o $ and
$Q_L+Q_R
= \sqrt{2} Q^e$.  Thus $Q^e$ measures the total charge on both edges
and is conserved even in the presence of the interaction.  Therefore,
the even boson decouples from the quantities we study (although it
does enter into the determination of the spectrum at the fixed points
\WA) and will play no role in the following. Moreover, the
backscattering current is the rate at which the charge
difference between right and left edges decreases (see below).
It is thus directly related to the odd charge by
 $\partial_t \Delta Q$.

The constraint of integrability is a very powerful one. As we will
see, it enables the calculation of exact thermal distribution
functions in an interacting theory. Not surprisingly, only certain
models are integrable. The impurity is also significant; only
particular types of impurity couplings preserve the integrability.
The Luttinger model without the impurity is well known to be
integrable, but it is easily shown that before the
non-local map \evenodd, the
impurity destroys the integrability.  However
it {\it is} integrable in the odd-boson basis.
The integrability  of the impurity interaction
 in the Luttinger model was first established
by \GZ\ in the
context of the boundary sine-Gordon model.
In order to make contact with this work, we
 map our theory involving the {\it massless}
left-moving odd boson $\phi^o$
on the line, with the point contact interaction at the
origin $x=0$, into a boundary sine-Gordon problem.
This is done using a standard  ``folding'' procedure.
{}From the  left-moving odd boson  $\phi^o (x+t)$
 on the full line, we define right {\it and} left moving
odd bosons on the half line $x >0$ by
$$
\phi^o_L(x,t) \equiv \phi^o ( x+t), \qquad x>0
$$
$$
\phi^o_R(x,t) \equiv \phi^o (- x+t), \qquad x>0
$$
This model is no longer chiral, but lives on the half-line.

There are many field theories on the
half-line which are integrable, the Kondo model being a
celebrated example \AFL.
Recently much effort has gone into
understanding their properties. In the following we exploit the
results of \GZ. The odd-boson model with interaction
becomes the massless limit $\Lambda \to 0$
of the massive sine-Gordon model on the half-line
\eqn\sg{H_{SG}={1\over 8\pi\nu}\int_0^l dx
\left[\left(\partial_x\phi^o\right)^2 + (\Pi^o)^2+
\Lambda\cos \sqrt{2}\phi^o\right] +
\lambda_1 \cos{1\over\sqrt{2}}\phi^o(0)}
[$\Pi^o$ is the canonical momentum conjugate
to the odd-boson field $\phi^o$.]
This model is integrable for any value of the
bulk mass $\Lambda$, and the  Luttinger
parameter $\nu$ \GZ. However, the factor of two ratio
in the argument of the bulk and boundary cosines
seems to be necessary for the integrability.
In the standard sine-Gordon conventions where the first
term has a $1/2$ in front, our
normalization corresponds to $\beta_{SG}^2=8\pi\nu$.
A lattice regularization of this field theory,
the XXZ spin chain with a $\sigma_x$ perturbation on
the boundary, was long ago shown to be integrable \MW.

Notice that we can ``fermionize'' to get back a
Luttinger model (for $\Lambda\ne 0$ we get the massive
Thirring model) on the half-line
for the odd-boson  degree of freedom. However, because of
the non-linear change of basis in \evenodd,
the properties of the odd-boson Luttinger model are not
the same as the original model. In particular, the original
Luttinger model is a free fermion when $\nu=1$,
while the odd-Luttinger fermion is free when $\nu=1/2$. This
shift arises because of the $\sqrt{2}$ in  \evenodd, which is
necessary to keep the same normalization of
the kinetic term. Thus an interacting fermion on the full line
can be mapped to a non-interacting one on the half-line.
This provides a simple way of understanding
the results of \Guin, where the $\nu=1/2$ model is mapped
onto two Ising models, one of which has a boundary magnetic field.
Two Ising models are well known to be equivalent to a free Dirac
fermion (up to boundary conditions), and it is easily seen
that $H_B$ corresponds to a boundary
magnetic field on one of them.

To recover our
massless model, we simply have to take the $\Lambda\to 0$ limit \FSW.
So we see that, after the few mappings described above, our impurity
problem is integrable as long as only one  of the impurity coupling
constants $\lambda_n$ in \delH\ is non-zero.
As noted before, this is natural for $1>\nu>1/4$, where
power counting shows that only the first coupling constant
$\lambda_1$  is relevant.

\newsec{The quasiparticles, their $S$ matrix and their distribution
functions}
\subsec{The quasiparticle spectrum}

We find the appropriate quasiparticle basis by studying the model
\sg\ at arbitrary $\Lambda$. As discussed carefully in \FSW,
the quasiparticle spectrum remains the same in
the massless limit $\Lambda \to 0$
of the massive sine-Gordon model on the half-line \sg.
Introducing the parameter $\Lambda$ and then
setting it to zero is not necessary for solving the problem, but
it gives an easy way of finding the basis where the quasiparticles
scatter off the boundary without particle production.

Integrability means that there is an infinite number of conserved
quantities which commute with each other and
 with the Hamiltonian, even in the presence
of the impurity \delH\ \GZ.
{}From this we see already that in
an integrable system a particular basis
of Hilbert space, in which all the infinite conserved
quantities as well as the hamiltonian
are {\it diagonal},  plays a special and simplifying role.
It is this basis on which we will focus.
This basis has a ``quasiparticle'' structure
similar to the ``particle''
Fock space of a non-interacting
theory. In particular, the eigenvalues
of the infinite conserved
quantities in this basis have the form
 $\sum_i p_i^n $, where $p_i$ are momenta of
individual quasiparticles, and where $n$ runs over an infinite subset
of the positive integers (each $n$ labels one conservation law).
In the Luttinger model these run over $n$ odd.
  The identification of $p_i$ as momenta of ``quasiparticles''
arises from the form of the eigenvalue of the
Hamiltonian in this basis, which is
$ H =v_F  \sum_i p_i $. This means that the  energies
of the ``quasiparticles'' are additive, justifying
the association of the particle concept with those
eigenstates.

These conservation laws have important
consequences for scattering.  The quasiparticles of this basis must
scatter off the impurity {\it without} particle production, i.e.\
one-by-one. This means that the scattering
matrix off the point contact is a product
of 1-body $S$ matrices, one for each quasiparticle.
  Away from the impurity (in the ``bulk''), the
quasiparticles scatter off of each other with a completely elastic
and factorizable two-body scattering matrix $S^{bulk}$.
This follows from an old  kinematic argument \refs{\Poly,\ZZ}\
which applies when there is any conservation law with $n>1$.
Factorizability means in the bulk that the $N$-body
bulk $S$ matrix is a product of 2-body $S$ matrices $S^{bulk}$.
Completely elastic means
that {\it individual} momenta are conserved in a collision:
all that can happen is that the momenta of the particles
get permuted. This does
not mean that bulk scattering is trivial; internal quantum numbers
can change, and even if scattering is diagonal (i.e.\ internal
quantum numbers do not change), the particles can have a
phase delay, i.e. the $S$ matrix may be a momentum-dependent  phase.

There are many bases for the Hilbert space of the
{\it massless} odd-boson theory,
which are related by not-necessarily-local mappings. For example,
plane waves obviously are eigenstates of $H_0$, but they are not
eigenstates of $H_0+ H_B$ \FSW. The basis of particles of \sg\
which are eigenstates
has been known for some time \ZZ. (Indeed, in a massive theory there
is only one particle basis; it is only for $\Lambda=0$ in \sg\ that
there is a choice.) At any
value of $\nu$, the spectrum contains a kink ($+$) and an antikink
($-$). These carry (odd) charges $ Q^o = 1/\sqrt{2}$ and
$-1/\sqrt{2}$, respectively.  Moreover, for $n-1<1/\nu\le n$, there
are $n-2 $ ``breather'' states which have no charge.  These breathers
exist in the regime where the fermion interaction in the
odd-Luttinger
model is attractive; in this language they correspond to
fermion-antifermion bound states.
These particles
span the Hilbert space of the left-moving odd boson; we label them by
indices $j,k,\dots$ running over the kink $(+)$, antikink $(-)$ and
breathers $(b)$. One can in fact check explicitly that that these
particles are the solutions of the classical limit of \sg\ for any
value of $\Lambda$ \FSW.

Henceforth, we set $\Lambda=0$ in \sg\ so that the
particles are massless.
We also ignore the even boson, since it
does not affect the current or conductance.
We will find more convenient to use the unfolded language, so
we will continue to discuss a purely left-moving theory on the full
line with an impurity.

Since the particles are massless,
a left mover has dispersion relation $E=-p$.
Instead of momentum, we use rapidity $\t$, which for a particle of
type $j$ is defined as
$$E=-p=m_je^{\t},$$
where $m_k =M\sin(k\pi\nu/2(1-\nu))$ for the $k$th breather
and $m_\pm=M/2$. The overall scale $M$ is arbitrary
and cancels out of all physical results.

Notice that the momenta of the quasiparticles
all have one sign  (the rapidity $\theta$ is real).
Since this might be unfamiliar, we now  express
the charged fermions occurring for $\nu =1/2$
in terms of kink and antikink quasiparticles; for
$\nu=1/2$ there is no breather. Recall that for $\nu=1/2$
we can re-fermionize the odd-boson theory. Thus we obtain
a single non-interacting,  spinless charge fermion
$\psi^{\dagger}(p),\psi(p)$
where the momentum $p$ runs over all real values (positive
and negative). This fermion satisfies canonical anticommutation
relations, since it is non-interacting. The kinks and antikinks
for this simple theory can be
defined using a canonical particle-hole transformation:
$$\eqalign{
\psi_+(p) &\equiv \psi(p) \qquad \qquad
\ \ \ (kink)\cr
\psi_-(p) &\equiv \psi^{\dagger}(-p) \quad
\qquad (antikink)\cr}
$$
for $p>0$.
The left-hand-side defines kink and antikink annihilation operators.
Similarly, the kink and antikink creation operators
are the hermitian conjugates:
$$\eqalign{
\psi_+^{\dagger}(p) &\equiv \psi^{\dagger}(p)
\qquad \qquad
(kink)\cr
\psi_-^{\dagger}(p) &\equiv \psi(-p) \quad
\qquad (antikink)\cr}
$$
for $p>0$.
Thus for $\nu=1/2$ kinks and antikinks are just particle-hole
transforms of ordinary fermions, and have only  one sign
of momentum. Unfortunately, for $\nu\ne 1/2$ the mapping is not
as straightforward.

While there are no statistics in $1+1$ dimensions (there is no way to
bring a particle around another without interacting), a crucial issue
is how many particles are are allowed to occupy each level. The
answer
is well established in the massive sine-Gordon model: only one
quasiparticle is allowed per level (they are like fermions). This is
basically because the Bethe-ansatz wave function used to solve the
model or its regularized lattice versions vanishes when two or more
excitations have the same rapidity. It is possible to study similarly
a regularized lattice version of the massless limit of the
sine-Gordon
model \refs{\FT,\RS} and to see that the same exclusion principle
holds in
that limit.  This does not contradict other ways of describing the
excitations, where the particles are free but obey ``exclusion
statistics''
\Haldexc. These particles seem to be related to the
quasiparticles of the Calogero- Sutherland model, or the
Haldane-Shastry spin chain. In the realization of the
Luttinger liquid in the fractional quantum Hall effect
discussed
in sect.\ 2, these particles are the Laughlin quasiparticles
restricted to the edge.  These particles are clearly not the same as
ours; for example they are the quasiparticles of the original
Luttinger impurity problem, while ours are those of the odd boson
theory, which is found from the original by a non-local map. This
means that there are (at least) two sets of bases of quasiparticles
for the Luttinger liquid, one where the particles are free but fill
levels in a peculiar manner, and another, where the particles fill
levels like fermions but interact. In the first approach, however, it
is not known yet how to include an impurity.  If this problem is
solved, it would be very interesting to compare the two methods of
calculation.

\subsec{The bulk and impurity $S$ matrices}

These massless kinks and breathers interact with each other and
with the  impurity. Even though the model away from the impurity
is a free boson, an effect of using this quasiparticle basis
is that the particles interact even in the bulk. We describe these
interactions with bulk and impurity scattering matrices.
First, we have an $S$ matrix for a particle scattering off of the
impurity.  Since the integrability requires that the particles
scatter one-by-one, this can be described by one-particle
$S$ matrix elements. Kinematically, all that happens is the
particle goes through the impurity with a phase delay.
Because the impurity interaction violates odd-charge
conjugation, it is possible for the positively-charged kink to
scatter
into a negatively-charged antikink when going through the
impurity.

Dimensional analysis tells us that the $S$
matrix elements must depend only on the ratio $p/T_B$.
If we define the ``backscattering rapidity'' by the relation
$$T_B={M\over 2} e^{\t_B},$$
this means that all impurity $S$ matrix elements depend on the
rapidity difference $\t-\t_B$.
Thus the impurity-kink $S$ matrix consists of the elements
$S_{++}(\t-\t_B)= S_{--}(\t-\t_B)$ for
 kink $\to$ kink, and antikink $\to$ antikink, as well
as $ S_{+-}(\t-\t_B)= S_{-+}(\t-\t_B) $ for
 kink $\to$ antikink, and vice versa.
These $S$ matrix elements are given by taking the massless
limit of the results of \GZ:
\eqn \bdrs {\eqalign{
S_{++}(\t)&={\exp\left(\l\t\right)
\over 1+ i\exp\left(\l\t\right)}
\exp[i\alpha_{\nu} (\t)] \cr
S_{+-}(\t)&={1\over 1+ i\exp\left(\l\t\right)}
\exp[i\alpha_{\nu} (\t)]
.}}
Here  $\exp[i \alpha_{\nu} ]$ is the phase of the expression
given in  Eq.(3.5)
of \FSW.
For convenience we have defined
$$\l\equiv{1\over\nu} -1. $$
The boundary $S$ matrix is
unitary: $|S_{++}|^2 + |S_{+-}|^2=1$.
The $S$ matrix is such that for particle with very large energy (UV
limit), the scattering is diagonal. Diagonal scattering in
the unfolded theory is totally off-diagonal scattering in the
folded
one, so matrix elements here are interchanged as compared to
\FSW.

Because we are no longer working in the plane-wave basis,
it is also necessary to find the non-trivial
$S$ matrix for particles interacting
in the bulk. Since all particles are massless, they must all have the
same velocity $v_F$ (which is set to 1 in this paper).
Thus it is not immediately obvious how to define
an $S$ matrix for two left movers. It is best interpreted as a
matching condition on the Bethe wave function, as we will explain in
the next subsection. Alternatively, one can think
of acting on multiparticle
states with creation and annihilation operators: the non-trivial $S$
matrix means that these operators satisfy non-trivial commutation
relations, the so-called Zamolodchikov-Faddeev algebra \ZZ.
Since there is no mass scale in the bulk (only $T_B$ at the
impurity),
a two-particle bulk $S$ matrix element can only depend on the ratio
of
the two momenta $p_1/p_2$. In terms of rapidity, this is a function
of
$\t_1-\t_2$.  Thus we can now think of the impurity as a particle
with
rapidity $\t_B$ and a different $S$ matrix than the bulk one.

For general $\nu$, the left-left two-particle $S$ matrix is given by
the same formula as in the massive case. This $S$ matrix is not
diagonal: the initial state $|K(\t_1)A(\t_2)\>$ can scatter to
$|A(\t_1)K(\t_2)\>$ because the kink $K$ and antikink $A$ have the
same mass.  For the kink/antikink scattering one has three
amplitudes \ZZ
\eqn\sll{\eqalign{a(\t)&=\sin[\l(\pi+i\t)]Z(\t)\cr
b(\t)&=-\sin(i\l\t) Z(\t)\cr
c(\t)&=\sin(\l\pi) Z(\t).\cr}}
where the $S$ matrix  element $a(\t_1-\t_2)$ describes the process
$|K(\t_1)K(\t_2)\> \rightarrow |K(\t_1)K(\t_2)\>$,
as well as
$|A(\t_1)A(\t_2)\> \rightarrow |A(\t_1)A(\t_2)\>$,
$b$ describes $KA\to KA$,
$c$ describes the
non-diagonal process $KA\to AK$, and there is a symmetry under
interchange of
kink to antikink (corresponding to $\phi^o\to -\phi^o$).
The
function
$Z(\t)$ is a normalization factor, which can be written as
$$Z(\t)={1\over \sin[\l(\pi +i\t)]}\exp\left(i
\int_{-\infty}^{\infty} {dy\over 2y}
\sin {2\t\l y \over\pi} {\sinh [(\l-1)y]
 \over \sinh y\cosh[\l y]} \right).$$
The breather-kink and
breather-breather $S$ matrices are well known
\ZZ; we do not write them down here.

When $1/\nu$ is an integer, the bulk scattering is
diagonal ($c$ vanishes) and $a=\pm b$.
Therefore, the only allowed
processes are of the form $ |j(\t_1)\> \otimes$ $|k(\t_2)\>$ $ \to
|k(\t_2)\> \otimes$ $ |j(\t_1)\>$. Such a process is described by the
$S$ matrix element $S^{bulk}_{jk}(\t_1-\t_2)$.
However,
the impurity scattering is not diagonal, so
charge transport is still possible.
The bulk diagonal scattering makes the thermodynamic Bethe ansatz
computation discussed in the next subsection much simpler at
these values of $\nu$.

\subsec{The non-fermi distribution functions}

As with the $S$ matrix, it is convenient to use rapidity
instead of momentum or energy to define the densities of states
and distribution functions.
The number of allowed kink or antikink states per unit length
with rapidity  between $\t$ and $\t+d\t$
 is given by $n_\pm(\t,V) d \t$,
while the number of states actually occupied by kinks or antikinks
in this rapidity range is
$n_+(\t,V) f_+(\t,V) d \t$ and $n_-(\t,V) f_-(\t,V) d \t$,
respectively.  Thus
\eqn  \DeltaQkink {
\<  \Delta Q\>_V = 2l \int_{-\infty}^\infty
d\t\ [n_+(\t,V)\ f_+(\t,V) - n_-(\t,V) f_-(\t,V)]
.}
Since at most one kink is allowed per level, we have
$0\le f_\pm \le 1$. The functions $n_j$ and $f_j$ are defined
for the breathers in the same manner.

These thermodynamic functions $n_j(\t)$ and $f_{\pm}(\t,V)$ are
different from the free-fermion functions when the odd-boson kink
theory is an interacting Luttinger liquid ($\nu\ne 1/2$), but we can
derive them exactly.  The idea is simple, and has become known as the
thermodynamic Bethe ansatz \Zamtba. It is basically the same as what
is
used in exact solutions of other impurity problems like the Kondo
problem \AFL. The main difference between the approach used here and
the traditional approach is that in the latter, the Bethe ansatz
appears as a tool to diagonalize a bare hamiltonian, while here we
work directly in the renormalized theory where factorized scattering
is assumed, and the Bethe-ansatz equations (here relations between
$n$
and $f$) follow simply from a matching condition on the
Bethe  wave function.
This matching condition gives one set of equations relating
the functions $n_j$ and $f_j$. Following the standard
thermal approach to Bethe ansatz \YY, one writes
the free energy as a functional
of $n$ and $f$ and minimizes it.
Together with the first equation from the matching
condition, this second set of equations from the minimization
requirement yields
$n$ and $f$.

The simplest situation occurs when the scattering is
completely diagonal,  which occurs when $1/\nu$ is an integer.
Here we can easily impose periodicity of the Bethe
wave function since we
know the $S$ matrix, which encodes the change of the wave function
when two particles (or a particle and the impurity) are interchanged.
Thus we know the phase the wave function picks up when a particle is
brought ``around the world'' through all the others.
We denote generically the kink, antikink and breathers by the index
$j$, so the two-particle
$S$ matrix elements are labeled
by $S_{jk}(\t_1-\t_2)$.
In the following calculation of the densities $n$ and
$f$, we can
ignore the phase shift due to scattering on the impurity
because it affects densities only to  order $1/l$, where
$l$ is the system size. This does not affect
the densities in the thermodynamic
limit $l \to \infty$, and will therefore
not show up in the computation of the conductance. (On the
other hand, however, in the absence of the voltage these $1/l$
effects on the densities can be computed and are responsible for
the impurity free energy \FSW.)
The requirement
of a periodic boundary condition under this operation quantizes the
momenta:
\eqn\quant{e^{-im_{r_i}\exp(\t_i)l/\hbar}\prod_{j=1,j\ne i}^{\cal
N}
S_{r_ir_j}(\t_i-\t_j)=1 \ ,}
where $E=-P= m_j \exp(\theta)$ for a particle of type $j$.
The periodicity condition \quant\ includes the interaction
of any particle with all the others.
By taking a
derivative with respect to the rapidity of the logarithm of
\quant, one finds
\eqn\ntof{
n_j(\t)= {m_j e^{\t} \over h}
+  \sum_k \int_{-\infty}^{\infty} d\t'
\Phi_{jk}(\t-\t') n_k(\t') f_k(\t'),}
where $\Phi_{jk}(\t)=-i (d / d \t) \ln S^{bulk}_{jk}(\t)/2\pi$ \YY.
For example, for $\nu =1/3 $ there is one breather $(b)$ and \KM
\eqn\phibb{\eqalign{
\Phi_{bb}(\t)&=~2\Phi_{++}(\t)=2\Phi_{+-}(\t)=
-{1\over \pi\cosh \t}\cr
\Phi_{b+}(\t)&=\Phi_{+b}(\t)=
-{\sqrt{2}\cosh\t\over \pi \cosh2\t},\cr}}
while the others follow from the symmetry $+\leftrightarrow -$.
Explicit expressions for $\Phi_{jk}(\theta)$ in the more
general case where $1/\nu$ is a positive integer are given
in Eq.(4.9) of \FSW, and in \KM.

One defines an auxiliary pseudoenergy variable $\epsilon_j$ to
parametrize $f_j$ via
$$f_j\equiv {1\over 1+e^{-\mu_j/T}e^{\epsilon_j}},$$
where the $\mu_j$ are the chemical potentials: $\mu_+=-\mu_-=eV/2$;
$\mu_b=0$.  By demanding that the free energy at temperature $T$
(expressible in terms of $f_j$ and $n_j$) be minimized, we find an
equation for $\epsilon_j $ in terms of the (known) bulk $S$ matrix
elements:
\eqn\tba{\epsilon_j(\t,V/T) = {m_j e^\t\over T} - \sum_k
\int_{-\infty}^{\infty} {d\t}\
 \Phi_{jk}(\t-\t')
\ln [1+ e^{\mu_k/T}e^{-\epsilon_k(\t,V/T)}].}
Solving this system of coupled integral equations for $\epsilon_j$
gives the functions $f_j$. Except in special cases, this solution
cannot be obtained in closed form, but it is easy to solve these
equations numerically. We will not need the
explicit $\Phi_{jk}$, because the equations \tba\ can
be simplified by using an trigonometric identity described
in \refs{\TS,\Zamo}\ for example. Denoting convolution
by $\star$
$$f\star g (\t) \equiv \int d\t' f(\t-\t')g(\t')$$
one finds
\eqn\incidence{
 \Phi_{jk}(\t) = \sum_l N_{kl} K\star\left(\Phi_{jl}(\t)
+ \delta_{jl}\delta (\t)\right)}
where  the kernel
$K(\t)\equiv {\l/(2\pi\cosh\l\t)}$, and $N_{jk}$ is the
incidence matrix of the following diagram

\bigskip
\noindent
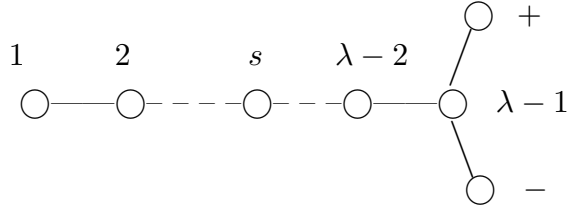
\centerline{\hbox{\rlap
{\raise28pt\hbox{$\hskip5.5cm\bigcirc\hskip.25cm
+$}}
\rlap{\lower27pt\hbox{$\hskip5.4cm\bigcirc\hskip.3cm -$}}
\rlap{\raise15pt\hbox{$\hskip5.1cm\Big/$}}
\rlap{\lower14pt\hbox{$\hskip5.0cm\Big\backslash$}}
\rlap{\raise15pt\hbox{$1\hskip1cm 2\hskip1.3cm s\hskip.8cm \l-2$}}
$\bigcirc$------$\bigcirc$-- -- --
--$\bigcirc$-- -- --$\bigcirc$------$\bigcirc$\hskip.3cm $\l-1$ }}

\bigskip
\noindent
i.e.\ $N_{jk}$=1 if the nodes $j$ and $k$ are connected, and $0$
otherwise (in particular $N_{jj}=0$).
This identity follows simply by Fourier transforming
the $\Phi_{jk}$ and using trigonometric identities.
We show this explicitly for $\nu=1/3$ ($\lambda=2$) in Appendix B.
Thus the simplified form of \tba\ is then
\eqn\tbaii{\epsilon_j(\t,V/T)= K\star\sum_k
N_{jk}\ln\left[1+e^{-\mu_k/T}e^{\epsilon_k(\t,V/T)}\right].}
The dependences on the mass ratios seems to have
disappeared from \tbaii, but they appear as an asymptotic condition:
the original equations \tba\ indicates that the solution must satisfy
$$\epsilon_j \to {m_j \over T} e^{\t}
\qquad \hbox{as}\ \t\to \infty.$$

We emphasize for later use that comparing the relations \tba\ and
\ntof\ gives
$$n_j (\t,V)=
{T \over h} \partial_\t \epsilon_j (\t,V).
$$
One effect of these equations is that the symmetry implies that
$$n_-(\t,V)=n_+(\t,V)\equiv n(\t,V)$$
 and
$\epsilon_-(\t,V)=\epsilon_+(\t,V)$.

The analysis at general values of $1/\nu$ is more complicated.
Since the bulk
$S$ matrix is not diagonal, the phase picked up when bringing a
particle around the world can be expressed only as an eigenvalue of a
monodromy matrix, which itself must be diagonalized. This can be
done at the price of introducing a further Bethe ansatz (see
\refs{\Zisn,\FI}\ for example). The
results are quite simple for $\nu/(1-\nu)$ integer: the equation
\tbaii\ holds, with the addition of a term $\delta_{j1}Me^\t /2$ to
the right-hand side. The general result is quite complicated, but
the appropriate diagonalization has been done for all $\nu$ \TS.
We caution that only for $1/\nu$ an
odd integer does this one-component Luttinger liquid apply to
the Hall effect; when $\nu $ is a more complicated
fraction, the edge states of the quantum Hall effect
 may be described by models with more than one boson \Wen.

\newsec{The current and the conductance}

We have shown that the bosonic field theory
with Hamiltonian \sg\ can be studied in terms
of a particular set of quasiparticles and their scattering.
We now compute an exact equation for the conductance
using this basis.

Without the backscattering,
the left and right charges
(or equivalently, the even and odd charges) are conserved
individually.
The backscattering allows processes where a charge carrier of the
left-moving edge hops to the right-moving edge or vice versa.
In the original basis, the current $I_B$ is the rate
at which  the charge of the left-moving edge is depleted.
By symmetry, $\partial_t Q_L = - \partial_t Q_R$ in each such hopping
event, so $I_B =  \partial_t  \bigl ( {e\over 2}\Delta Q \bigr ) =
\partial_t  \bigl ( {e\over \sqrt{2}} Q^o \bigr ),$
and we see that in the even/odd basis, the tunneling
corresponds to the violation of odd charge conservation
at the contact. In the $S$ matrix language
this happens when $S_{+-}\ne 0$, so that
a particle of positive odd charge
(the kink) can scatter into one of negative charge (the antikink) at
the contact. Neutral quasiparticles
cannot transport charge and thus do not
directly contribute to $\partial_t \Delta Q$.
We emphasize that neither the bulk nor the boundary
$S$ matrix elements depend on the voltage; the voltage only
affects the thermodynamic properties.

To calculate the conductance, we start with a gas of quasiparticles
with a chemical potential difference for kinks and antikinks
corresponding to the voltage $V$.  A positive voltage means that
there
are more kinks.  When there are more kinks than antikinks, the
backscattering will turn more kinks to antikinks than it turns
antikinks to kinks.  When a kink of momentum $p$ is scattered into an
antikink (the conservation laws require that it have the same
momentum
$p$) this changes $\Delta Q$ by $-2$.  Since kink and antikink
quasiparticles scatter off the point contact one-by-one, we may
describe the rate at which this charge transport occurs in terms of
two quantities: the probabilities of finding a kink or antikink of
momentum $p$ at the contact, and the transition probability
$|S_{+-}(\t-\t_B)|^2$.  We therefore study the density of states
$ n(\t,V) \equiv n_+(\t,V)
 =n_-(\t,V)$ and the distribution functions $f_\pm(\t,V)$ in
the thermodynamic limit ($l \to \infty$) and in the presence of an
applied voltage $V$.  We can now compute the backscattering current
from a rate (Boltzmann) equation.  The number density of kinks of
rapidity $\t$ which scatter into antikinks per unit time is given by
$|S_{+-}|^2 n(\t,V) \rho_{+-}$, where $\rho_{+-}$ is the probability
that the initial kink state of rapidity $\t$ is filled {\it and} the
final antikink state is empty. In a fermi liquid, we would have
simply
$\rho_{+-}= f_{+}[1-f_{-}]$.
However, in our interacting theory, correlations
between the particles mean that this does
not necessarily factorize in this manner. We can however write
$\rho_{+-}=f_+ - f_{+-}$, where $f_{+-}$ is the (unknown)
probability that the kink and antikink state are {\it both occupied}.
The rate at which antikinks scatter to
kinks is likewise proportional to $\rho_{-+}=f_- - f_{+-}$. Thus the
charge $Q_L-Q_R$ changes
at a rate proportional to $\rho_{+-}-\rho_{-+}=
f_+ - f_-$, independent of the unknown factor.
Using \DeltaQkink\ with $I_B=e\del_t Q^o/\sqrt{2}$ gives
\eqn \Boltzmann {
I_B(T_B,V,T)=
-e\int_{-\infty}^\infty d\t\ n(\t) |S_{+-} (\t-\t_B)|^2
\left[ f_+(\t,V) -
 f_-(\t,V) \right].}
Without the backscattering, the current is $I_0(V)
=e \<\Delta Q\>/2l$, and it follows from
\DeltaQkink\ that $I_0$ is given
by the same expression without the $-|S_{+-}|^2$. Thus the full
current
is proportional to $1-|S_{+-}|^2=|S_{++}|^2$.
Using the definition of $f$ and the result $n (\t)=
{T \over h} \partial_\t \epsilon_\pm (\t)$ allows us
to simplify the resulting expression. Notice that
$$n f_\pm= -{T \over h} {\partial\epsilon_\pm\over \partial\t}
{\partial \over \partial \epsilon_\pm}
\ln(1+ e^{\pm eV/2T}e^{-\epsilon_\pm})=
-{T \over h} {\partial \over \partial \t}\ln(1+ e^{\pm
eV/2T}e^{-\epsilon_\pm})$$
Defining
$$\epsilon(\t,V)\equiv \epsilon_+(\t-\ln(M/2T),V)=
\epsilon_-(\t-\ln(M/2T),V)$$
and plugging into \Boltzmann\ gives
our main result
\eqn\curri{\eqalign{I(T_B,V,T)
&= {eT\over h}  \int_{-\infty}^{\infty} {d \t}\
{1\over 1+ e^{2\l(\t-\t_B)}}
\del_\t \ln \left[{1 + e^{eV/2T}
e^{-\epsilon(\t+\ln(M/2T),V)}\over
1 + e^{-eV/2T}e^{-\epsilon(\t+\ln(M/2T),V)}}\right]
 \cr
&= {eT\l \over 2h}  \int_{-\infty}^{\infty} {d \t}\
{1\over \cosh^2\left[\l(\t-\ln(T_B/T))\right]}
\ln \left[{1 + e^{eV/2T}e^{-\epsilon(\t,V)}\over
1 + e^{-eV/2T}e^{-\epsilon(\t,V)}}\right]
 \cr
}}
where to get to the second line we integrated by parts and
redefined $\t$ by a shift. Even though the breathers
do not appear in \curri,
they interact with the kink and antikink and
affect the calculation of $\ep$.

The differential
conductance is defined as $G(T_B/T,V/T)=\del_V I(T_B,V,T)$. In the
$V\to 0$ limit, the result can be written in a simple form.
Using \tba\ and
\ntof, it is easy to see that $d\epsilon/dV|_{V=0}=0$, so \FLSi
\eqn\condeq{G(T_B/T,0)=
{e^2\l\over 2h}  \int_{-\infty}^{\infty} {d \t}\
{1 \over 1 + e^{\epsilon(\t,0)}}
 {1\over \cosh^2\left[\l(\t-\ln(T_B/T))\right]}}

To check our result, we consider $\nu=1/2$, where
the conductance was previously derived exactly \KF. As noted above,
the odd-boson kinks are simply free fermions \refs{\Guin, \FSW,
\ALxray},
so they have the Fermi distribution
function $f_\pm(\t,V)=1/[1+
\exp((Me^{\t}\mp eV)/2T)]$ implying
that the function $\epsilon (\t)$ is
simply $\epsilon(\t)=e^{\t}$. This of course follows from \tba,
because at $\nu=1/2$ the bulk scattering is trivial and
$\Phi_{jk}=0$.
Note again, as mentioned  in Section 3.1 above, that
we have used the kink/antikink description
of the free fermion theory here, where all momenta have
only {\it one} sign. Using the particle-hole transformation
of Section 3.1, one finds immediately that
the occupation number of kinks and antikinks
are
$$ f_{\pm }(p) = \Theta(p) { 1\over 1 + e^{(p \mp  V/2)/T}}, \qquad p
>0,\ \nu=\half$$
whereas the occupation number of unoccupied kink/antikink
states are
$$[ 1-f_{\pm }](p) = \Theta(p) { 1\over 1 + e^{(- p \pm V/2)/T}},
\qquad p >0,\ \nu=\half$$
Now all momenta have one sign. Note that
$f_{\pm} (p=0,V=0) = 1/2$.
These fermions  (alias kinks) do
scatter non-trivially off of the point contact,
with $S$ matrix given  by \bdrs.
The resulting expression for $G(T_B,0)$ obtained from \condeq\  is
identical to the result in sect.\ VIII of \KF. Actually one can
re-express
the integral in terms of dilogarithm functions after some lengthy but
straightforward manipulations to find \Weissi
\eqn\exactonehalf{I(T_B,V,\nu=\half)={e^2V\over 2h}\left[1-{2T_B\over
eV}Im\psi\left({1\over 2}
+{T_B\over 2\pi T}+{ieV\over 4\pi T}\right)\right],}
where the digamma function $\psi(x)=\Gamma'(x)/\Gamma(x)$.
At zero voltage in particular one finds
\eqn\exactonehalfi{G(T_B,0,\nu=\half)={e^2\over 2h}\left[1-{T_B\over
2 \pi
T}\psi'\left({1\over 2}+{T_B\over 2\pi T}\right)\right],}
and at zero temperature
\eqn\exacthalfii{I(V,\nu=\half)={e^2V\over 2h}-{e T_B\over
h}\arctan{eV\over
2 T_B}.}
These $\nu=1/2$ formulas also apply for $\nu$ near
$1/2$ in the leading-logarithm approximation,
after a $V$- and $T$-dependent renormalization
of $T_B$ \Weiss.

Only at $T=0$ can these equations be solved in closed form for
all $T_B$; we discuss this limit in the next subsection.
However, we can study the solutions in certain limits.
As $T_B/T\to 0 $, we can
evaluate the conductance explicitly,
The linear-response conductance \condeq\
becomes
$$G(0,0)=[f_\pm(-\infty,0)-f_\pm(\infty,0)]e^2/h$$ in this limit.
One finds $f_\pm(\infty,V)=0$
obviously from \tba\ in this limit.
More generally, one has
\eqn\curruv{I(T_B=0,V,T)=-{eT\over h}
\ln\left[1-(1-e^{-eV/T})f_+(-\infty,V) \right].}
To find $f(0,V)$, we use a
well-known trick, given for example in \Zamtba.
The kernels in \tba\ or \tbaii\ are appreciably
different from zero only when $\t'$ is near $\t$.
Thus when we are interested in values of $\t$ near $-\infty$,
we can replace the value of $\epsilon(\t',V)$ in the integral
with $\epsilon(-\infty,V)$. We then
can do the integral over the kernel explicitly.
This then gives us the coupled difference equations
$$\epsilon_j(-\infty,V)=  \sum_k {N_{jk}\over 2}
\ln (1 + e^{-\mu_k/T}e^{\epsilon_k(-\infty,V)})$$
Solving these explicitly gives $f(-\infty,0)=\nu$
and we indeed recover $G_0=\nu e^2/h$.  For the non-equilibrium
conductance, it is not even obvious that $G_0$ will not depend on
$V$. However, one can check that
$$f_+(-\infty,V)=\exp\left({eV(1-\nu)\over 2T}\right){\sinh\nu
eV/2T\over \sinh eV/2T}$$
and plugging this into \curruv\ gives $G_0=\nu e^2/h$ for all $V$.

These non-perturbative equations also give the
perturbative exponents.
As $T_B/T\to \infty$, the linear-response conductance
$G\propto(T/T_B)^{2(1-\nu)/\nu}$.  Thus it goes to
zero with the correct exponent, as in \KF.
For $T_B/T$
small, it can be argued that
$\epsilon(\t+i\pi/(1-\nu))=\epsilon(\t)$
\Zamo. Using this to write a power series for
$\epsilon$ and plugging into \curri\  gives $G- G_0
\propto
(T_B/T)^{2(1-\nu)}$. Both are in agreement with \KF.  In fact, when
$\nu <1/2$ all of the coefficients $g_n(V/T)$ in the series
$G=\sum_{n=0}^\infty g_{2n} (T_B/T)^{2n(1-\nu)}$ can be computed
using
Jack polynomial technology \FLSjack.
Moreover, there is a non-perturbative functional relation \FLSjack\
relating the linear-response
conductance to the free energy for all $\nu <1$.

We also note that there has been some confusion about the power of
the exponent for $T/T_B$ near zero. As discussed
in \FSW, the leading (irrelevant) perturbation is the
energy-momentum tensor, which is of
scaling dimension two. This in fact means that the leading correction
to the free energy is of order $T^2$. (This power
was also derived by scaling arguments in \Guineanew.) However, this
operator
does not give any contribution to the DC
conductance. Intuitively, this
is because the energy-momentum tensor has no charge and should not
affect charge transport. More precisely, one can check that when
inserted
into the Kubo formula, powers of the frequency
$\omega$ appear. When we take the DC
$\omega\to 0$ limit this contribution vanishes. Thus the naive
scaling
arguments of \Guineanew\ do not apply to the DC linear-response
conductance. However, outside of linear response
at $V>0$, one finds indeed that
$$I(T_B,V,T)\approx I(T_B,V,0)+ T^2 I_2(T_B,V) +...$$
in agreement with \refs{\Weiss,\Guineanew}.

We can easily solve for the conductance numerically.
To plot the complete function, one fixes a value
of $V/T$ and solves \tba\ numerically for $\epsilon$ to
double-precision
accuracy and inserts the
result into \curri.  Evaluating the integral numerically
for various values of $T/T_B$ then gives
$I(T_B/T,V/T)/V$ as a function of $T_B$ for fixed $V$ and $T$.
To find the conductance
it is easiest to just vary the voltage slightly in order to take the
derivative numerically. A more precise way would be to use the
fact that once one knows $\epsilon$ numerically, the integral
equations
for $d\epsilon/dV$ are linear and can be solved by inverting
large matrices (of size the number of lattice sites used in the
discretization of the integral) numerically.

Several graphs of $G$ for $\nu=1/3$ are given in the figures.
For $V/T < 8$, the plot is qualitatively the same as for $V=0$:
a flat region at $G=e^2/3h$ for $T$ large, a transition region where
the power series corrections cause it to fall off until
it reaches its asymptotic form proportional to
$(T/T_B)^{4}$.
However, at $V/T\approx 8$, a qualitatively new
feature appears: $G$ has a peak! It is not very sharp;
the highest it gets for $\nu=1/3$ is $G\approx .35 e^2/h$.
A variety of values of $V/T$ are plotted in fig.\ 1 and
fig.\ 2. In the first we plot it versus $T_B/V$ in order
to make the approach to the $T=0$ limit to be discussed in sect. 6
clear. In the second we plot it versus $T_B/T$; we see
that there is a substantial broadening of the curve as the voltage
is increased. This should provide a prominent signal in the
experiments.

This peak and the values of $V/T$ for which it occurs can be
understood theoretically.  A peak occurs for voltages large enough so
that $g_2(V/T)$ changes sign.  This must happen because at zero
temperature the current can be expanded for large voltage in
the form
$$I(V)\propto V(\nu + C \left(T_B \over
V\right)^{2(1-\nu)} + \dots)$$
Because the backscattering must make the current decrease we have
$C<0$. Taking the derivative with respect to $V$ we see that
for $\nu < 1/2$ the conductance must increase for small
enough $V/T_B$.  We can even find the value $V^*$ where
$g_2(V^*/T)=0$
analytically, because $g_2$ can be calculated in
closed form \FLSjack.
This is done by first calculating the coefficient $Z_2$
of the partition function in the case where the interaction (in
Euclidean time) is $\cos[\sqrt{2}\phi(0,\tau)+ 2\pi p
\tau T]$, which is \FLSjack $$Z_2(p)={\sin{\pi\nu}\
\Gamma(1-{2\nu})\over
\sin\pi (\nu+p)\Gamma(1-\nu+p)\Gamma(1-\nu-p)}.$$
The finite voltage case is obtained by analytically continuing $2\pi
p$ to $i\nu eV/T$ \KF. Then we use the relation $g_2 (V/T) \propto Re
\del _p Z_2(p)$ (which can be shown by using explicit perturbation
theory in the impurity interaction) to give
\eqn\gtwo{g_2(V/T)\propto Re
\left[Z_2\left({i\nu eV\over 2\pi T}\right)
\{\psi(\nu+{i\nu eV\over 2\pi T})-
\psi(1-\nu+{i\nu eV\over 2\pi T})\}\right].}
Solving for $g_2(V^*/T)=0$ gives $eV^*/T=7.18868564374998...$ for
$\nu=1/3$ and $eV^*/T=6.653022289582846...$ for $\nu=1/4$.

The perturbative results of \FLSjack\ give a
completely independent check on the TBA results.
One can fit the TBA  results to a power series numerically
to determine the perturbative coefficients to compare; one
indeed finds for example that $g_2(V^*/T)=0$ in the TBA results.
One can also check that the relation $T_B=C\lambda_1^{1/1-\nu}$
is independent of $V$, showing that the boundary $S$ matrix is indeed
independent of voltage.

\newsec{Explicit Solution at $T$=0}

Some remarkable simplifications take place in the $T=0$ limit.
Although the densities of states are still non-trivial, the
distribution functions $f$ become step functions. As a result, the
TBA
equations become linear and can be solved using the Wiener-Hopf
technique. We find explicit series expressions for $I(V/T_B)$ and
$G(V/T_B)$ in this limit. Moreover, this leads to an exact duality
between large $V/T_B$ and small $V/T_B$. In the Hall devices this
corresponds to a duality between Laughlin-quasiparticle tunneling
and electron tunneling.

At $T=0$ and $V=0$ the ground state of the theory is just the vacuum
with neither kinks nor breathers; these particles are in fact
defined as excitations above this vacuum.
When $V$ is turned on, this ground
state becomes unstable. For $V>0$, kinks of charge $e$ start
filling the vacuum, since they are energetically favorable for small
enough momentum (large negative rapidity).  The new ground state is
made of kinks occupying the range $\theta\in (-\infty, A]$;
in other words, $f_+(\t,V)=1$ for $\t<A$ and $f_+(\t,V)=0$
for $\t>A$.
The surface of the sea is approximately
$A \approx \ln (eV/M)$,  but
computing $A$ exactly
requires some technology because the kink interaction affects the
filling of the sea.
There are no antikinks nor breathers in the sea at $T=0$,
so their densities do not appear in this analysis.
For ease of
notation we define $\rho(\t)\equiv n_+(\t,V)f_+(\t,V)|_{T=0}$.
When $T=0$ the
periodicity relation \ntof\ reduces to the following coupling between
kink rapidities:
\eqn\dens{2\pi n_+(\t)={M\over 2\hbar}e^\theta
+2\pi\int_{-\infty}^A
\Phi(\theta-\theta')\rho(\theta')d\theta',}
where
$\rho=0 \hbox{ in }[A,\infty)$, and $\Phi=\del_\t S_{++}(\t)/2\pi$
follows from the kink-kink bulk $S$ matrix \sll.

We consider now the general equation
\eqn\geneq{\rho(\theta)-\int_{-\infty}^A
\Phi(\theta-\theta')\rho(\theta')d\theta'=g(\theta),\quad \theta\in
(-\infty,A],}
where in the above example $g(\theta)={M\over 2h}e^\theta$ for
$\theta\in
(-\infty,A]$, $g(\theta)=0$ otherwise. By taking Fourier
transforms, and since $\rho$ vanishes outside $(-\infty, A]$ we have
\eqn\eqi{\int_{-\infty}^\infty d\omega
e^{-i\omega\theta}\left\{\tilde{\rho}(\omega)[1-\tilde{\Phi}(\omega)]
-\tilde{g}(\omega)\right\}=0,\quad \theta\in (-\infty,A],}
where we defined Fourier transforms by
$$\tilde{h}(\omega)=
\int h(\theta)e^{i\omega\theta}d\theta,\quad
h(\theta)=\int\tilde{h}(\omega)
e^{-i\omega\theta}{d\omega\over 2\pi}.$$
For any function $h$ we also introduce
$$h_-(\omega)\equiv \tilde{h}(\omega) e^{-i\omega A}.$$
This subscript should not be confused with the antikink index used
in other sections of this paper.

The integral equation \geneq\ for $\rho$ can now be written in the
form
\eqn\eqii{\tilde{\rho}(1-\tilde{\Phi})-\tilde{g}=e^{i\omega
A}X_+(\omega),}
where $X_+(\omega)$ is analytic in the upper half plane.
Since $\rho$ vanishes in $[A,\infty)$ we see that $\rho_-$
is analytic in the
lower half plane, and similarly for $g_-$. To implement
the Wiener-Hopf technique (see the appendix in \JNW\ for a very clear
exposition),
we need to factorize $1-\tilde{\Phi}$
into the form
\eqn\eqiv{1-\tilde{\Phi}={\sinh[(\l+1)\pi\omega/2\l]\over
2\cosh[\pi\omega/2]
\sinh[\pi\omega/2\l]}\equiv {1\over G_+G_-},}
where $G_+(G_-)$ is analytic in the upper (resp. lower) half plane
and  $G_+(G_-)$ vanishes only in the lower (resp. upper) half plane.
One finds
\eqn\eqv{G_+=\sqrt{2\pi(\l+1)}
{\Gamma(-i(1+\l)\omega/2\l)\over
\Gamma(-i\omega/2\l)\Gamma(1/2-i\omega/2)} e^{-i\omega\Delta},}
$$G_-(\omega)=G_+(-\omega),$$
$$\Delta\equiv{1\over 2}\ln \l-{1+\l\over 2\l}\ln(1+\l) .$$
The phase ensures analyticity of $G_+$ at $i\infty$.
Having set up the equation in this form, we can write
down the solution for $\rho$:
\eqn\eqviii{{\rho_-\over G_-}=[g_-G_+]_-,}
where
$$[h]_-(\omega)\equiv-{1\over 2i\pi}\int_{-\infty}^\infty
{h(\omega')\over  \omega'-\omega+i0}d\omega'.$$
In our case
$$\tilde{g}(\omega)={M\over 2h}
{e^{(i\omega+1)A}\over i\omega+1},$$
so we find
\eqn\eqxii{\tilde{\rho}(\omega)={M\over 2i h}
{G_-(\omega)G_+(i)\over \omega-i}e^{(i\omega+1)A}.}
This relation was obtained also in \Tsv.

We have not yet determined the dependence of the cut-off $A$
on the physical
potential $V$. This can be done in two ways.
The first way consists in minimizing the energy of the system.
Following
\FSZII\ we introduce a function $\epsilon(\theta)$ which satisfies
\eqn\eqxv{{eV\over 2}-{M\over 2}e^\theta=\epsilon(\theta)-
\int_{-\infty}^A\Phi(\theta-\theta')\epsilon(\theta')d\theta',}
and the cut-off follows now from the condition $\epsilon(A)=0$, which
reads in terms of fourier transforms
$$\lim_{\omega\rightarrow\infty}
\omega\epsilon_-(i\omega)=0.$$
We can easily solve for $\epsilon$. This is the same equation as the
one for
$\rho$ except for the substitution
$$g(\theta)={eV\over 2 }-{M\over 2}e^\theta,\quad
\tilde{g}(\omega)=\pi eV\delta(\omega)
-{M\over 2}{e^{(i\omega+1) A}\over i\omega+1}.$$
We find therefore
\eqn\eqxvii{\epsilon_-(\omega)=-{M\over 2i}{G_-(\omega)G_+(i)\over
\omega-i}e^{A}+{eV\over 2i}{G_-(\omega)G_+(0)\over \omega}.}
It follows that
\eqn\eqxviii{e^A={eV\over M}{G_+(0)\over G_+(i)}.}
%
%and thus
%
%\eqn\eqxix{\tilde{\rho}(\omega)=
%{eV\over 2h}{G_-(\omega)G_+(0)\over i\omega+1}e^{i\omega A}.}
%

The second method of determining $A$ is to study the energy
in the absence of backscattering, which is
$$
\eqalign{E&= \int_{-\infty}^A d\t\, \rho(\t)
\left[{M\over 2}e^\t - {eV\over 2}\right]
={M\over 2}\tilde\rho(-i) - {eV\over 2}\tilde\rho(0)\cr
&={M^2\over 8h}G_+(i)G_-(-i)e^{2A} - {eM\over 4h}G_-(0)G_+(i)
Ve^A.\cr}
$$
This energy results from the massless
description obtained by considering the free field
as  the $\lambda\to 0$ limit of the sine-Gordon model \sg.
We can recover this energy directly from the Hamiltonian
in this limit. Coupling
to a potential amounts to considering the Hamiltonian
(in the original left-right basis)
\eqn\eqxxi{H_0={\hbar\over 4\pi\nu} \int_{-l}^l dx
\left[(\partial_x\varphi_R)^2 + (\partial_x\varphi_L)^2\right] -
{eV\over 4\pi}\int_{-l}^l dx
\left[\partial_x\varphi_L +\partial_x\varphi_R \right].}
Redefining the fields $\phi_L$ and $\phi_R$ by a shift
removes the linear term,  and gives the energy
per unit length associated with the potential $V$.
Equating the two energies gives $A$ in agreement
with formula \eqxviii.

Since we know the density explicitly from \eqxii\ and \eqxviii,
an explicit form for the current follows from
\Boltzmann, which at $T=0$ reads
\eqn\current{\eqalign{I(V,T_B)&=e\int_{-\infty}^A \rho(\theta)
\left|S_{++}\left(\t-\t_B\right)\right|^2 d\theta \cr
&={e^2 V\over 2h}\int_{-\infty}^0
d\theta\ F(\theta)
{1\over 1+ \exp[-2\l(\t+\ln(eV/T'_B)-\Delta)]} .\cr}}
where
$$\tilde{F}(\omega)\equiv {G_-(\omega)G_+(0)\over 1+i\omega},$$
and the boundary temperature is redefined:
$$T'_B\equiv T_B 2 e^{-\Delta}{G_+(i)\over G_+(0)}=
T_B {2\sqrt{\pi}(\lambda+1)\Gamma({1\over 2} + {1\over 2 \lambda})
\over \Gamma({1\over 2 \lambda})}.$$

We can therefore extract the following results from \current.
At very large potential, the $S$ matrix element goes to one and
$I_0=G_0V$, with
\eqn\asympcond{
G_0= {e^2\over 2h} \int_{-\infty}^0 F(\theta)d\theta={e^2\nu\over
h}.}
For small voltage, we can expand the $S$ matrix element
in powers of $V/T'_B$ to get
\eqn\iexpans{\eqalign{I(V,T_B)=&{e^2V\over 2h}\int_{-\infty}^0
F(\theta)d\theta
\sum_{n=1}^\infty (-)^{n+1}\left({eV\over T'_B}
e^{\theta}\right)^{2n\l}\cr
\equiv&{e^2V\over 2h}\sum _{n=1}^\infty I_{2n}
\left({eV\over T'_Be^{\Delta}}\right)^{2n\l},\cr}}
where
$$I_{2n}={(-1)^{n+1}\over 2\pi} \int_{-\infty}^\infty
{G_-(\omega)G_+(0)\over (1+i\omega)(2n\l-i\omega)}d\omega=
(-1)^{n+1} {G_-[-2in\l]G_+(0)\over 1+2n\l}.$$
This expansion is valid only for $eVe^{-\Delta}/T'_B<1$.
By using \eqv, we find our final result for the low-voltage
expansion
\eqn\expbis{I={e^2\over h}V
\sum_{n=1}^\infty (-1)^{n+1}
{\sqrt{\pi}\,
\Gamma({n\over \nu})\over 2\Gamma(n)\Gamma({3\over 2}+n({1\over
\nu}-1))}
\left({eV\over T'_B}\right)^{2n({1\over \nu}-1)},\qquad
{eV\over T'_B e^{\Delta}}<1.}

We prove in the appendix that this strong-barrier expansion
can be transformed into the following weak-barrier expansion
\eqn\expter{\eqalign{I={e^2V\nu\over h}\left[1-
\sum_{n=1}^\infty
(-1)^{n+1}
{\nu\sqrt{\pi}\Gamma({n\nu})\over 2\Gamma(n)\Gamma({3\over 2}+
n(\nu-1))}
\left({eV\over T'_B}\right)^{2n(\nu-1)}\right],
\quad {eV\over T'_B e^{\Delta}}>1.\cr}}
Up to a constant and a shift, this is the same expression as
for strong backscattering,
with the replacement $\nu\to {1\over \nu}$. We therefore have the
result
\eqn\duality{I(T'_B,V,\nu)={e^2\nu V\over h}-{\nu^2}
I(T'_B,V,{1\over \nu}).}
(In this equation, $T_B'$ is treated as a constant --- it is
the same on both sides, although we have seen $T_B'/T_B$
depends on $\nu$.)
This proves that for $T=0$ at least, the weak-barrier (small $T_B$)
and strong-barrier (large $T_B$) are completely dual to each other.
In the Hall device, this duality is a concrete proof of
a long-suspected
relation between electrons and Laughlin quasiparticles.
This arises from the fact that for weak backscattering,
the operator for the tunneling of Laughlin quasiparticles is
relevant while the electron tunneling operator
is always irrelevant \refs{\Wen, \KF}. Laughlin quasiparticles are
allowed to tunnel because they are tunneling through the bulk of the
sample. However, in the strong backscattering limit, only electrons
can tunnel because they are not tunneling through the Hall fluid.
Thus
the least irrelevant operator arises from electron tunneling. As seen
in \KF, the associated exponents are indeed related by duality; we
find
here that the entire expansions around these two limits are related.

This duality was somehow known in the literature \Sch,  but its
status is not totally clear to us. It is usually considered as only
approximate
since  it relies on an instanton
approximation in the large-barrier limit \KF. However, one can prove
an exact duality \FZ\ (in the context of  dissipative quantum
mechanics, which is equivalent) between the cosine problem in the UV
and the ``tight-binding'' problem in the
zero-temperature limit, and then in \GHM\ the
tight-binding problem is mapped back onto the cosine problem,
providing a sort of proof. The main source of difficulty is that
the zero-temperature
action must be handled with great care. Perturbation theory
around the zero-temperature fixed point
is ill-behaved and depends on an infinite number of counterterms
(see the discussion at the end of \Zisn\ in the case of the flow from
tricritical Ising to Ising model), so identifying the leading term in
the approach to this fixed point is not sufficient. This is
equivalent to saying that what one calls  the strong-barrier problem
must  actually  be defined with great care. The strong-barrier
problem which is at the end of our renormalization-group
 trajectory follows formally
from  dimensional continuation of the integrals for the weak barrier
problem. It is not in any case a generic strong-barrier problem.
For instance regularizing the
integrals with a short-distance cut-off would give very different
results,
with a non-monotonic conductance \FZ. Also, note
that the duality does not apply to the impurity free energy, for
example
as explained in \FSW\ for small $T$ the leading contribution is
proportional to $T^2$ for all $\nu$.

Interestingly \expter\  can be
compared with the perturbative expansion in \KF\
\eqn\pertcomp{I(V)={e^2\nu V\over h}
\left[1-{\pi^2\nu \over \Gamma(2\nu)}
\left(
{eV\nu}\right)^{2({\nu}-1)}(2\l_1\kappa^{-\nu})^2\right],}
thus providing the relation between the parameter $\lambda_1$ in the
action
and
the TBA parameter $T_B$:
\eqn\relal{\lambda_1\kappa^{-\nu}={2^{\nu}\over 4\pi}\Gamma(\nu)
\left(\nu T_B' \right)^{1-\nu}.}
where $\kappa$ is the non-universal (dimensionful) cutoff which
appears in the boson two-point function, as defined in \FLSjack.

Finally, setting $\nu={1\over 2}$ (i.e.\
 $\lambda=1$) in these formulas
gives  $G_+(in)=1$ so
$I_n={(-1)^{n+1}\over 2n+1}$ and $T'_B=4 T_B$. Hence
$$
I(V)={e^2V\over 2h} \sum_{n=1}^\infty
{(-1)^{n+1}\over 2n+1}\left({eV\over T_B}\right)^{2n}=
{e^2V\over 2h}-{eT_B\over h}\arctan {eV\over
2 T_B},
$$
in agreement with the $T\to 0$ limit of eq \exactonehalf.

\vfill\eject

\leftline{\bf Acknowledgements:}

We thank the many participants of the conference SMQFT
(May '94, USC)
 for discussions, and more particularly I. Affleck, M.P.A. Fisher,
F.D.M. Haldane, N.P. Warner and E. Wong.
We also thank D. Freed, C. Mak and F.P. Milliken for valuable input.
This work was supported by the Packard Foundation, the
National Young Investigator program (NSF-PHY-9357207) and
the DOE (DE-FG03-84ER40168), as well as by a Sloan Foundation
Fellowship (A.W.W.L).

\appendix{A}{Duality}

In this appendix we provide details of the derivation of duality
between
strong- and weak-tunneling regimes.
Defining
$$
b=\left({T'_B e^{\Delta}\over eV}\right)^{2({1\over \nu}-1)},
$$
we see that there are two different regions
to consider in the computation of the
integral
$$
\int_{-\infty }^0 {e^{-i\omega\theta}\over 1+be^{-2({1\over\nu}
-1)\theta}}
$$
depending on whether $be^{-2({1\over \nu}-1)\theta}$ is greater or
smaller than one. In each region, one can expand the integrand
appropriately, and integrate term by
term. For $V/T_B$ small, this piece is always smaller than one,
and the expansion \iexpans\ follows immediately.
For $V/T_B$ large, one divides the integral
in \current\ into
two pieces to get
\eqn\monstereq{\eqalign{
I(V)=&{e^2V\over 2h}PP\int_{-\infty}^\infty {d\omega\over
2\pi}{G_-(\omega)G_+(0)
\over 1+i\omega}\left\{
\sum_{N=0}^\infty (-1)^{N+1}
{b^{2N}\over i\omega
+2({1\over\nu} -1)N}\right.\cr
&\left. +b^{-i\omega/(2({1\over\nu}-1))}\sum_{N=0}^\infty
\left[{(-1)^N\over
i\omega +2({1\over\nu}-1)N}
+{(-1)^N\over -i\omega+2({1\over\nu}-1)(N+1)}
\right]\right\}.\cr}}
The poles of the rational functions
except the one for $\omega=0$ are all cancelled out
because $G_-(2N({1\over\nu}-1)i)=0$ for $N>0$.
For the first sum we close the contour in the lower half plane.
Since $G_-$ is regular in this regime,
the only pole is at $\omega=0$.
For the second sum
we close the contour in the upper half plane where there are
poles of $G_-(\omega)$ at
$$
\omega=2in(\nu-1),\quad n>0.
$$
as well as those of the rational function at $\omega=0$.
Collecting all terms gives
\eqn\monstereqbis{\eqalign{
I(V)=&{e^2V\over 2h}\left\{{2\nu}+4\sqrt{\pi}(\nu-1)
\sum_{n>0}(-1)^n{1\over 1+2n(\nu-1)}
{1\over \Gamma({-n\nu})
\Gamma(1/2+n(\nu-1))}\right.\cr
&
\left.{1\over 2({1\over\nu}-1)}
\left({T'_B\over eV}\right)^{2n(1-\nu)}
\sum_{N=0}^\infty \left[{(-1)^N\over N-{n\nu}}+{(-1)^N
\over (N+1)+{n\nu}}\right]\right\}.\cr}}
Using the identity
$$
\sum_{N=0}^\infty {(-1)^N\over N-{n\nu}}
+{(-1)^N\over N+1+{n\nu}}=
-{\pi\over\sin \pi {n\nu}}
$$
together with standard gamma-function identities gives \expter.

\appendix{B}{Incidence Matrix}

In this appendix we give some details involving the manipulations
of the ``incidence matrix'' of section 4, for the
case $\nu=1/3$.  We start by writing
%\eqn\bsbulk1{
$$
{\bf \Phi}(\theta)\equiv
\pmatrix{
\Phi_{++} &\Phi_{+b} &\Phi_{+-} \cr
\Phi_{b+} &\Phi_{bb} &\Phi_{b-} \cr
\Phi_{-+} &\Phi_{-b} &\Phi_{--} \cr} (\theta)
$$
%}
Using the explicit expressions given in
\phibb, one finds  that this may be written in
the form
%\eqn\BSbulk2{
$$
{\bf \Phi}(\theta)\equiv
\Phi_{++}(\theta) {\bf N}^2 + \Phi_{+b}(\theta) {\bf N}
$$
%}
where
$${\bf N} \equiv \pmatrix{ 0&1&0\cr 1&0&1\cr 0&1&0\cr} $$
It is convenient to fourier transform the rapidity dependence,
in order to turn convolutions into multiplications.
We find
%\eqn\bfourier1{
$$
 \tilde\Phi_{++}(k) =\int_{-\infty}^{\infty} d \theta  e^{ik\theta}
\Phi_{++}(\theta)
=-{1\over 2 \cosh (k \pi/2)}
$$
% }

%\eqn\bfourier2{
$$
 \tilde\Phi_{+b}(k) =\int_{-\infty}^{\infty} d \theta  e^{ik\theta}
\Phi_{+b}(\theta)
=- { \cosh(k \pi/4)\over \cosh (k \pi/2)}
$$
% }

Making use of the relation $ {\bf N}^3 = 2 {\bf N}$,
we find in fourier space
$$
{\bf N} \cdot \tilde{\bf \Phi}(k)=
\tilde\Phi_{+b}(k)
 {\bf N}^2 + 2 \tilde\Phi_{++}(k) {\bf N}
$$
If  we define the multiplicative factor
$$\tilde K(k) =  \tilde\Phi_{++}(k)/ \tilde\Phi_{+b}(k) =
{1 \over 2\cosh(k \pi/4)}$$
we find
$$
\tilde K {\bf N} \cdot \tilde{\bf \Phi}=
 \tilde{\bf \Phi} + (2 K \tilde\Phi_{++} - \tilde\Phi_{+b} ) {\bf N}
$$
The expression in parenthesis equals
$(2 \tilde K \tilde\Phi_{++} -
\tilde\Phi_{+b} )  = 1/( 2 \cosh(k \pi/4))=K(k)$.
In conclusion we have
found
\eqn\concl{
\tilde K {\bf N} \cdot \tilde{\bf \Phi}=
 \tilde{\bf \Phi} + \tilde K {\bf N}.}
Fourier-transforming back to rapidity space, we find
the relation  \incidence\ of section 4.
Note that for $\nu=1/3$,
$$
K(\theta) =\int_{-\infty}^{\infty}  { d k\over 2 \pi}
e^{-ik\theta} \tilde K(k)= {1 \over \pi \cosh(2 \theta)}
$$
as indicated below \incidence.

\listrefs
\vfill\eject
\centerline{\bf Figure Captions}
\bigskip
\noindent Fig.\ 1 .
The non-equilibrium conductance as a function of $\log(T_B/V)$,
for various values of $V/T$. Notice that
the curve develops a peak
when $V/T > 7.18868$, and that the $T \to 0 $ limit is smooth.
\bigskip
\noindent Fig.\ 2.
The non-equilibrium conductance as a function of $T_B/T$ for
various values of $V/T$. The curve broadens substantially
as $V/T$ is increased, even before developing a peak.
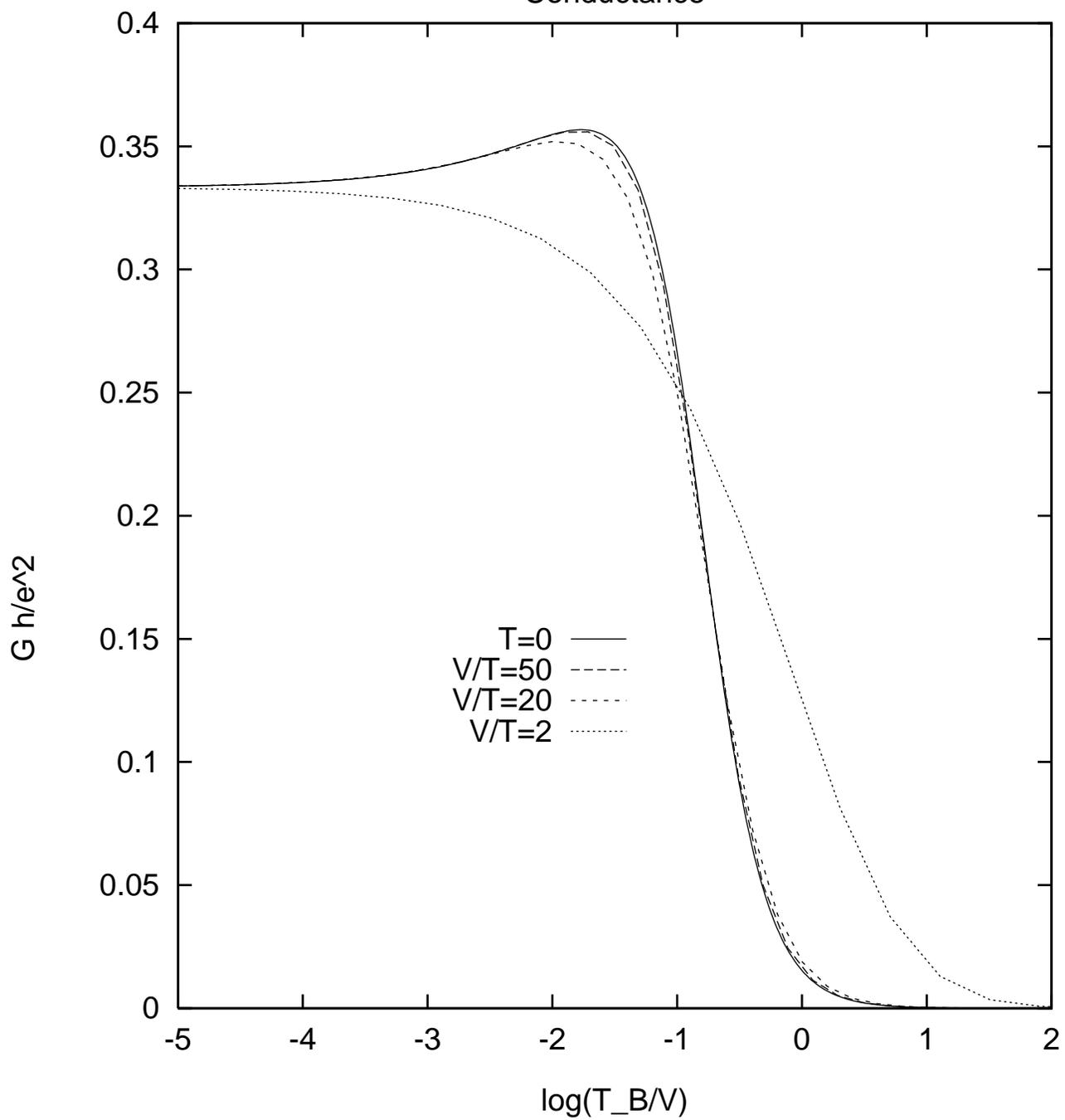
\bye